\documentclass[10pt, letter, onecolumn]{arxiv}

\usepackage{kantlipsum, lipsum}
\usepackage{dm-colors}
\usepackage{amsmath}
\usepackage{pstricks, pst-node}
\usepackage{verbatim}
\usepackage{multirow}
\usepackage{scalerel}
\usepackage{booktabs}
\usepackage{enumitem}
\usepackage{xspace}
\usepackage{bm}
\usepackage{bbm}
\usepackage{mathtools}
\usepackage{soul}
\usepackage{epsfig}
\usepackage{graphicx}
\usepackage{tcolorbox}
\tcbuselibrary{skins}
\usepackage{subcaption}
\usepackage{amssymb}
\usepackage{colortbl}
\usepackage{csquotes}
\usepackage{etex}
\usepackage{setspace}
\usepackage{svg}
\usepackage{colortbl}
\usepackage{tabularx,ragged2e}
\usepackage{placeins}
\usepackage[symbol]{footmisc}
\usepackage[bibstyle=nature,citestyle=numeric-comp,%
            natbib=true,backend=biber,maxbibnames=99,%
            giveninits=false,sorting=none]{biblatex}
\usepackage{nameref}
\usepackage{varioref}
\usepackage[pagebackref=false,breaklinks=false,%
            colorlinks=true,bookmarks=true,citecolor=ourdarkblue,%
            urlcolor=ourdarkblue,linkcolor=ourdarkblue]{hyperref}
\usepackage[noabbrev,capitalize]{cleveref}
\usepackage{etoc}

\graphicspath{{figures/}}

\title{{Towards Democratization 
\\ of Subspeciality Medical Expertise}}

\author[3,$\ast$]{Jack W. O'Sullivan}
\author[1,$\ast$]{Anil Palepu}
\author[2]{\\Khaled Saab}
\author[1]{Wei-Hung Weng}
\author[2]{Yong Cheng}
\author[3]{Emily Chu}
\author[3]{Yaanik Desai}
\author[3]{Aly Elezaby}
\author[3]{\\Daniel Seung Kim}
\author[3]{Roy Lan}
\author[3]{Wilson Tang}
\author[3]{Natalie Tapaskar}
\author[3]{Victoria Parikh}
\author[3]{Sneha S. Jain}
\author[2]{\\Kavita Kulkarni}
\author[2]{Philip Mansfield}
\author[1]{Dale Webster}
\author[1]{Juraj Gottweis}
\author[2]{Joelle Barral}
\author[1]{\\Mike Schaekermann}
\author[2]{Ryutaro Tanno}
\author[2]{S. Sara Mahdavi}
\author[1]{Vivek Natarajan}
\author[1]{Alan Karthikesalingam}
\author[3,$\dagger$]{\\Euan Ashley}
\author[2,$\dagger$]{Tao Tu}

\affil[1]{Google Research, }
\affil[2]{Google DeepMind, }
\affil[3]{Stanford University}

\renewcommand{\correspondingauthor}[1]{$\ast$~Equal contributions. %
                                       $\dagger$~Equal leadership. \\%
                                       $\ddagger$~Corresponding authors: \{jackos, euan\}@stanford.edu, \{natviv, taotu\}@google.com}

\begin{document}

\begin{refsection}

\begin{abstract}

The scarcity of subspecialist medical expertise, particularly in rare, complex and life-threatening diseases, poses a significant challenge for healthcare delivery. This issue is particularly acute in cardiology where timely, accurate management determines outcomes. We explored the potential of AMIE (Articulate Medical Intelligence Explorer), a large language model (LLM)-based experimental AI system optimized for diagnostic dialogue, to potentially augment and support clinical decision-making in this challenging context. We curated a real-world dataset of 204 complex cases from a subspecialist cardiology practice, including results for electrocardiograms, echocardiograms, cardiac MRI, genetic tests, and cardiopulmonary stress tests. We developed a ten-domain evaluation rubric used by subspecialists to evaluate the quality of diagnosis and clinical management plans. Evaluation was blinded to whether plans were produced by general cardiologists or AMIE, the latter enhanced with web-search and self-critique capabilities.

AMIE was rated superior to general cardiologists for 5 of the 10 domains (with preference ranging from 9\% to 20\%), and equivalent for the rest. Access to AMIE's response improved cardiologists' overall response quality in 63.7\% of cases while lowering quality in just 3.4\%. Cardiologists' responses with access to AMIE were superior to cardiologist responses without access to AMIE for all 10 domains. Qualitative examinations suggest AMIE and general cardiologist could complement each other, with AMIE thorough and sensitive, while general cardiologist concise and specific - analogous to the combination of a sensitive initial diagnostic test (AMIE) followed by a highly specific confirmatory test (general cardiologist). Overall, our results suggest that specialized medical LLMs such as AMIE have the potential to augment general cardiologists' capabilities by bridging gaps in subspecialty expertise, though further research and validation are essential for wide clinical utility.

\end{abstract}

\maketitle


\section{Introduction}
\label{sec:Main}
Globally, there is a significant shortage of speciality medical expertise~\citep{WHO2016}. The World Health Organization (WHO) predicts a deficit of 18 million providers by 2030, with shortages being most acute in resource-limited and rural areas~\cite{charlton2015challenges}. This disparity is exacerbated for rarer and more complex conditions, particularly those for which timely treatment prevents morbidity and mortality. For instance, Hypertrophic Cardiomyopathy (HCM) is one of the leading causes of sudden cardiac death in young adults~\citep{Ommen2024-em}, yet, more than half of US states do not have a HCM subspecialist center~\citep{hcmcenter}. Lack of subspecialist access has led to 60\% of HCM patients undiagnosed in the US~\citep{Massera2023-og}. With premature mortality highly preventable with implanted cardiac defibrillators (ICDs)~\citep{Ommen2024-em}, cardiac conditions such as HCM exemplify this urgent unmet need in healthcare delivery: timely and widely-available access to subspecialist expertise~\citep{Maddox2024-ci}. While cardiac conditions serve as an indicative example, the consequences of delayed access to subspecialist care are significant across all specialities, often resulting in increased morbidity and mortality as patients miss critical diagnostic and treatment windows. Navigating the cascade of referrals required to access subspecialist expertise creates undue stress and anxiety while presenting a time-consuming and resource-intensive process for both patients and healthcare providers.

Large language models (LLMs) have emerged as potential assistive tools for an array of healthcare issues~\citep{Clusmann2023-cf, van2024adapted}. LLMs can rapidly synthesize data from multiple sources and suggest differential diagnoses and management plans~\citep{tu2024towards, mcduff2023towards, saab2024capabilities}. Notably, some LLMs have already been integrated into electronic medical record software~\citep{Jennings2024, turner2023epic} as assistive tools for summarization and communication. Despite the potential of LLMs to enhance medical expertise, rigorous assessment of their performance remains scarce in medical specialities, with few openly-available datasets for model evaluation. It remains unclear whether LLMs possess the nuanced understanding and intricate knowledge base required to effectively replicate the decision-making process of experts in highly specialized medical fields~\citep{hager2024evaluation}.

This study probes the potential of LLMs to democratize subspecialist-level expertise by focusing on an indicative example, the domain of genetic cardiomyopathies like HCM. Our key contributions are as follows:
\begin{itemize}
\item \textbf{A novel dataset for cardiovascular disease:} We introduce and open-source a dataset encompassing cardiac testing and genetic information from 204 real-world patients at the Stanford Center for Inherited Cardiovascular Disease (SCICD), enabling further research in this specialized field.
\item \textbf{Leveraging AI for specialized cardiovascular assessments:} We utilize Articulate Medical Intelligence Explorer (AMIE)~\cite{tu2024towards}, an LLM optimized for diagnostic dialogue, to generate detailed assessments of patients with suspected complex cardiovascular disease.
\item \textbf{Benchmarking AI and general cardiologists:} We propose a ten-axis rubric that subspecialists utilize to evaluate the quality of diagnosis, triage and clinical management proposals for complex cardiology cases. Under blinded evaluation, subspecialists compare AMIE's performance to that of general cardiologists and demonstrate AMIE's assessments are preferred in 5 out of 10 tested domains (and equivalent in the rest), indicating its potential to provide assistive expertise. However, AMIE also exhibited a higher rate of clinically significant errors.
\item \textbf{Augmenting cardiologist expertise with AI:} We explore the potential of LLMs to up-level general cardiologists by evaluating whether access to AMIE's response improves their clinical decision-making in this specialized area. We observe that after seeing AMIE's response, the general cardiologist assessments are preferred over their unassisted response for all 10 domains when evaluated blindedly by subspecialists.
\item \textbf{Illustrating clinical applications:} We qualitatively demonstrate, by simulating scenarios based on four patients in our dataset, the potential clinical applications of AMIE in conversing with patients and assisting general cardiologists.
\end{itemize}

\begin{figure}[htp!]
    \centering
    \includegraphics[width=\textwidth]{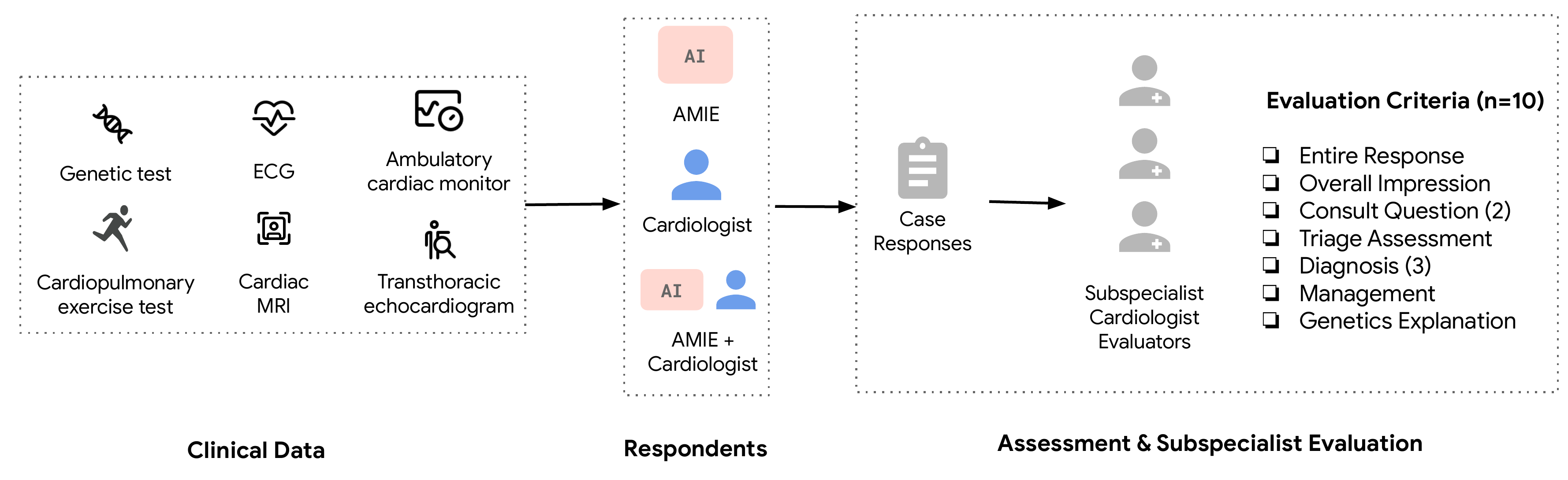}%
    \caption{\textbf{Study design.} Text reports from the cardiac testing data of 204 patients with suspected genetic cardiovascular disease were provided to AMIE as well as general cardiologists. AMIE and the cardiologists each answered the assessment form listed in \cref{fig:Assessment}. Later, the cardiologists were allowed to view AMIE's responses and make any changes to their initial assessments. Subspecialist cardiologists from the Stanford Center for Inherited Cardiovascular Disease provided individual ratings as well as direct preferences between AMIE and cardiologists (\cref{fig:direct}) and between the cardiologist responses before and after seeing AMIE's assessment (\cref{fig:assist}). Subspecialists were blinded to the source of ratings, and reviewed responses in a randomised sequence.}%
    \label{fig:studydesign}%
\end{figure}

\clearpage
\section{Methods}
\label{sec:Methods}

To assess AMIE's ability, we developed a blinded fully-counterbalanced reader study, as described in \cref{fig:studydesign}. The study comprised four main phases: 1. Acquisition of clinical data: Recruitment and de-identification of clinical data from a subspecialized inherited cardiovascular center, 2. Domain Adaptation of AMIE (Articulate Medical Intelligence Explorer), 3. General cardiologist and AMIE assessment of each case, followed by revised assessment of the cases by general cardiologists when given access to the AMIE assessment outputs, and 4. Subspecialist evaluation and analysis. General cardiologists and AMIE were tasked with interpreting clinical text data from 204 real-world patients, including electrocardiograms (ECGs), rest and stress transthoracic echocardiograms (TTEs), genetic testing, cardiac MRIs (CMRs), ambulatory Holter monitors, and cardiopulmonary exercise tests (CPX). To facilitate scientific progress and reproducibility of our results, we have made all data publicly available (see \textit{Data Availability}). 

This data was obtained from patients referred to the Stanford Center for Inherited Cardiovascular Disease, encompassing patients with both suspected and confirmed inherited cardiovascular diseases, and general cardiology patients. We recruited three board-certified general cardiologists not affiliated with Stanford, who have not had dedicated training in cardiovascular genetics and had not cared for any of the studied patients. These general cardiologists and AMIE were asked to diagnose, triage, and manage these patients using the provided text reports. The general cardiologists assessed the data twice: initially without AMIE support and later with assistance from AMIE's assessment. Afterwards, blinded subspecialist cardiologists used a multi-domain evaluation rubric to compare the responses from AMIE to those from general cardiologists, as well as the general cardiologist's responses before and after assistance from AMIE.

\subsection{Clinical data}

The Stanford Center for Inherited Cardiovascular Disease (SCICD) is one of the world’s largest centers focused on inherited cardiovascular disease. Patients with suspected cardiovascular disease are referred for diagnosis and management. Referrals typically are from general cardiologists. The patient population largely contains patients with suspected inherited cardiovascular disease, however the clinic also sees general cardiology patients. All physicians at SCICD have received subspeciality training in genetic cardiac disease.

The assessment of patients involves review of the patient’s history and review of tests such as cardiac MRIs, rest and stress echocardiograms, cardiopulmonary stress tests, ECGs, ambulatory Holter monitors, and genetic testing. For each of these tests, a text report outlining in-depth and summarized test results is produced. Text results from these investigations were available and utilized for all patients included in this study. We utilized 204 real-world patient cases in the test set after first exposing the model to 9 different patient cases for model refinement. No cases utilized in model refinement were utilized for model testing.

\subsection{AMIE: An LLM based AI system for diagnostic dialogue}
\subsubsection{Model development}

\begin{figure}[htp!]
    \centering
    \includegraphics[width=\textwidth]{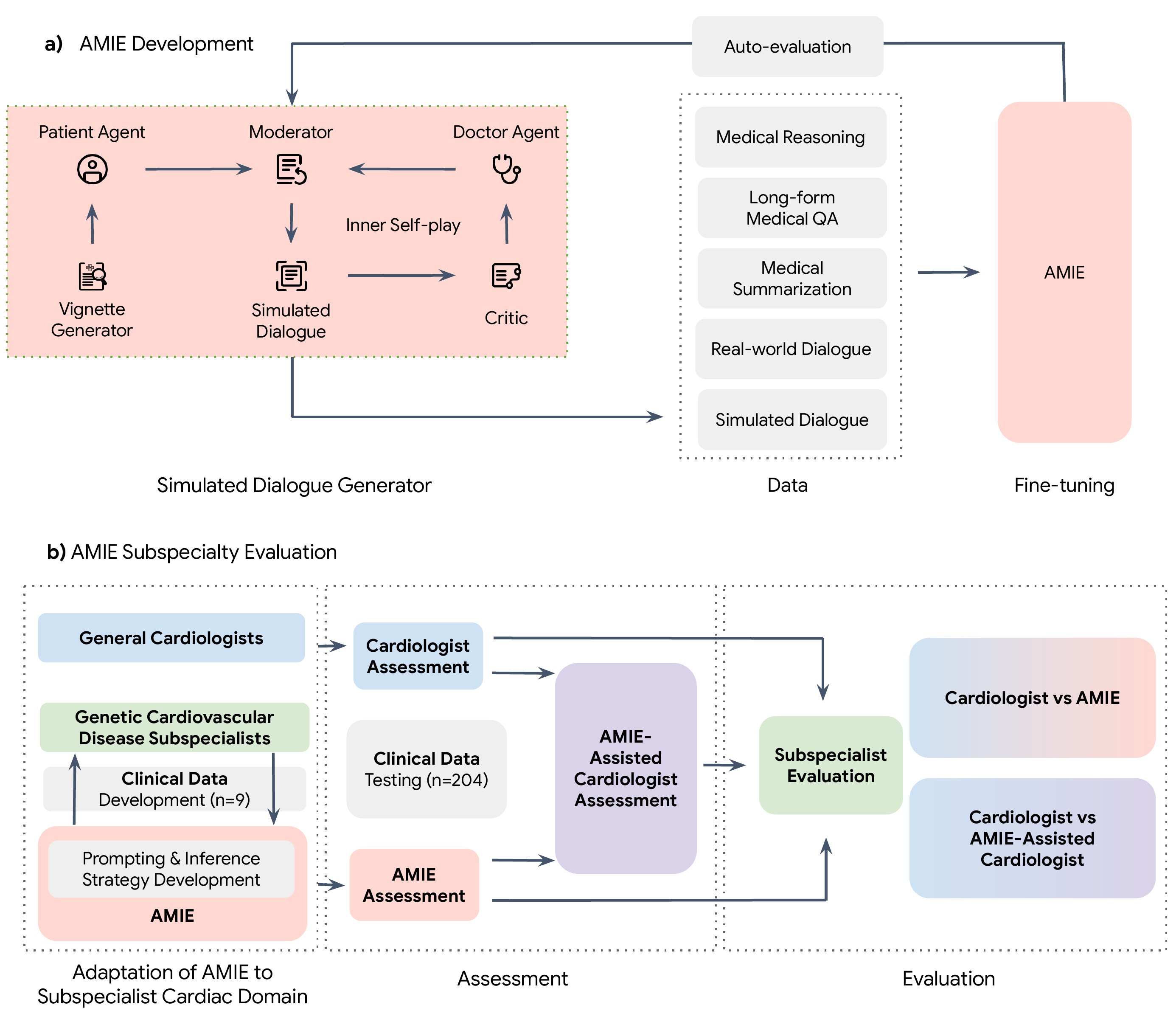}%
    \vspace{0.2cm}
    \caption{\textbf{a) Development of AMIE.} AMIE was trained with a self-play based simulated learning environment (see \cite{tu2024towards} for details). We leveraged AMIE without any additional instruction fine-tuning. \textbf{b) Specialization and evaluation of AMIE.} Of the 213 total cases, 9 were used to iterate on the prompting and inference strategy, while the rest were used to test AMIE and the cardiologists. During the study, after individually completing the assessment form in \cref{fig:Assessment}, cardiologists could see AMIE's response and alter their response. Subspecialist cardiologists from the Stanford Center for Inherited Cardiovascular Disease provided individual ratings (\cref{fig:Box2}) and direct preferences (\cref{fig:Box3}) between AMIE and cardiologists, and between the cardiologist responses with and without assistance from AMIE.}%
    \label{fig:amie}%
\end{figure}

Articulate Medical Intelligence Explorer (AMIE), is an experimental LLM-based medical artificial intelligence (AI) system optimized for clinical history-taking and diagnostic dialogue~\cite{tu2024towards}. AMIE was trained with a self-play based simulated learning environment for diagnostic medical dialogues, enabling the scaling of AMIE’s knowledge and capabilities across a multitude of medical conditions and contexts (see \cref{fig:amie}). Specifically, we previously used this environment to iteratively fine-tune AMIE with a corpus of medical question answering, reasoning, summarization tasks and real-world medical dialogue data in addition to an evolving set of simulated dialogues. At inference time, AMIE utilizes a chain-of-reasoning strategy adaptable to various medical tasks, enabling the integration of automated feedback mechanisms and tool use to refine its response.

\subsubsection{Domain adaptation of AMIE}
Adaptation of AMIE to this subspecialist domain required clinical data from just nine patients. Five cases, chosen at random, were used to design an optimal approach to model prompting through iterative review and expert feedback. Several prompting strategies were tested: zero vs. few-shot prompting, with vs. without search retrieval augmentation, providing an AMIE generated summary of the cardiac testing information vs. providing raw cardiac testing information, and using the original response vs. allowing the model to self-critique and revise its answer. After AMIE produced responses on the 5 patients using various combinations of these strategies, we had a specialist cardiologist review the responses and select their preferred set of responses. Ultimately, the specialist preferred the ``few-shot with self-critique and search retrieval augmentation'' method, in which AMIE drafts an initial response to the case with few-shot exemplars, conducts web search to retrieve relevant guidelines, and then critiques and revises its initial drafted response. The few-shot exemplars used were four additional patient cases for which a specialist cardiologist provided a ``gold-standard'' case assessment. Note that none of the nine patient cases used as either few-shot exemplars or for validating inference strategies were part of the test set.

\subsection{Study design and evaluation}
\subsubsection{Cardiologist and AMIE assessment}

We exposed both AMIE and board-certified general cardiologists without subspecialty expertise in genetic cardiovascular disease to test results from 204 consecutive, real-world patients. This test population consisted of patients suspected or confirmed to have inherited cardiovascular disease, as well as a mixture of patients without genetic cardiovascular disease. Test results included: cardiac MRIs, rest and exercise echocardiograms, cardiopulmonary stress tests, ECGs, and ambulatory Holter monitor results. Genetic test result data were also available for some of the patients. 

Both AMIE and general cardiologists completed the same standardized assessment form shown in~\cref{fig:Assessment}. After AMIE and the general cardiologists completed their assessments individually, the cardiologists underwent a washout period of 2 months. Then, the cardiologists were again provided with the clinical data, as well as AMIE's and their own responses, and asked to revise their assessment if needed.

The assessment form (\cref{fig:Assessment}) aims to evaluate the respondents across a range of domains including triage, diagnosis, and management of these patients with potential inherited cardiovascular disease. Respondents provide an `Overall impression' of the patient's case and answer a `Consult Question' regarding the likelihood of a genetic cardiomyopathy.  The `Triage Assessment' section prompts respondents to determine the necessity of referral to a specialist center. In the `Diagnosis' section, the respondent is asked to list their most likely diagnosis as well as any additional information they would need to ask the patient or gather from tests. In the `Management' section, the respondent describes their management plan and any additional information they would need. After answering these previous sections, the respondent is provided with genetic test results (where available) and asked if and how they would change any prior answers.

\begin{figure}[htbp!]
\begin{tcolorbox}[
    colback=black!5!white,
    colframe=black!60!white,
    title=\textbf{Cardiologist/AMIE Assessment Form},
    fonttitle=\bfseries,
    arc=3mm,
    boxrule=1pt,
    bottomrule=2pt,
]
\footnotesize
        
\textbf{Overall impression:}\newline This is a free text response to summarize overall assessment of the patient’s test results.\newline 
\textbf{Consult question:}\newline
Does this patient have a genetic heart disease? Please provide a brief rationale.\newline 
\textbf{Triage assessment:}\newline
Does this patient require a referral to a specialist center for genetic cardiovascular disease?\newline 
a) No referral required: low likelihood of inherited cardiovascular disease.\newline
b) Referral required to inherited cardiovascular disease center: High suspicion for inherited cardiovascular disease, refer to specialist.\newline
c) More information required.\newline
\textbf{Diagnosis:}\newline 
1. What is the most likely diagnosis (Free text response)? \newline 2. Did you have enough information to make a diagnosis (Yes/No)?\newline 3. What further information would you need to ask the patient? \newline 4. What further information from tests would you need to aid the diagnosis? \newline  
\textbf{Management:}\newline
1. What is your proposed management plan?\newline
2. Did you have enough information to make a management plan (Yes/No)? \newline
3. What further information would you need to ask the patient to aid the management? \newline
4. What further information from tests would you need to aid the management?\newline
\textbf{Genetic test results:}\newline
After reviewing the genetic test results (where available), do the genetic test results change any of your responses? 
\end{tcolorbox}
\vspace{0.2cm}
\caption{\textbf{Assessment Form for AMIE/cardiologist responses to cases.} AMIE and cardiologists were provided clinical text from various cardiac testings for each patient and asked to complete the assessment form. They initially completed all but the last question without genetic test results, and then were provided any available genetic test results to answer the last question.}
\label{fig:Assessment}
\end{figure}

\subsubsection{Subspecialist evaluation}

We performed a series of comparisons to examine the utility of AMIE to appropriately triage, diagnosis, and manage patients with suspected inherited cardiovascular disease: 
\begin{enumerate}
  \item General cardiologist response vs. AMIE response;
  \item Individual assessment of AMIE and general cardiologist responses;
  \item General cardiologist response vs. general cardiologist response with AMIE assistance. 
\end{enumerate}

The responses of 204 real-world cases were evaluated by four subspecialist cardiologists, who were blinded to source of responses. We conducted two types of evaluation: direct A/B preference comparisons and an individual assessments, described further in \cref{sec:rubric}. 

The statistical analysis of the results were performed via bootstrapping (see \cref{sec:stats}).

\subsubsection{Development of an evaluation rubric for subspecialist case interpretation}
\label{sec:rubric}
We developed rubrics for subspecialists to evaluate responses, under two conditions: Direct A/B comparison of general cardiologist's and AMIE's responses for each domain, and individual evaluation of specific responses in isolation.

For the direct A/B comparison (see \cref{fig:Box3}), the domains of the evaluation rubric mirrored the clinical assessment form completed by general cardiologists and AMIE. For each domain, evaluators indicated a direct preference between general cardiologists and AMIE with an option for a tie. We chose this direct comparison with a third option for a tie to facilitate greater potential discrimination in performance between general cardiologists and AMIE compared with Likert scales, while also allowing equivalence to be expressed.  To investigate more nuanced qualities of each response, subspecialists also evaluated general cardiologist and AMIE responses individually. To do this we developed an individual evaluation rubric (see \cref{fig:Box2}).
This was developed through identification of response quality themes from the existing literature \citep{van2024adapted}, followed by an iterative combination of semi-structured feedback from LLM domain experts and cardiologist experts not involved in the evaluation or the remainder of the study design. Themes from the existing literature were shared with respective experts. Experts then provided feedback in a semi-structured manner. Themes were narrowed and refined in response to expert feedback until thematic saturation (no new themes emerged) was reached, which in our case lead to five major themes. Instructions for evaluators were written and piloted using the 9 development cases, these cases were piloted and used as worked examples in interactive feedback sessions with expert evaluators. Feedback on instructions, and the evaluation rubric was sought from experts and changes implemented if concordance from experts was present. The above approach led to iterative improvement to our individual evaluation rubric, spanning five crucial domains of LLM evaluation: 1. Errors, 2. Addition, 3. Omission, 4. Reasoning and intelligence, and 5. Bias. Subspecialist experts were asked to answer Yes or No to direct questions spanning these 5 domains and then given free text responses to quantify and explain their selections. 

\begin{figure}[h!]
\begin{tcolorbox}[
    colback=black!5!white,
    colframe=black!60!white,
    title=\textbf{Subspecialist Preference Evaluation Form},
    fonttitle=\bfseries,
    arc=3mm,
    boxrule=1pt,
    bottomrule=2pt,
]
\footnotesize
For each of the following criteria, indicate which response you prefer? \textbf{Options: Response 1, Tie, Response 2}
\begin{enumerate}
    \item The entire response
    \item Overall Impression
    \item Consult question: Does this patient have a genetic heart condition?
    \item Consult question answer: Brief explanation of your answer
    \item Triage assessment: Does this patient need a referral to a center specializing in genetic heart disease?
    \item Diagnosis: What is the most likely diagnosis?
    \item Diagnosis: What further information would you need to ask the patient to aid the diagnosis?
    \item Diagnosis: What further information from tests would you need to aid the diagnosis?
    \item Management
    \item Do the genetic test results change your response?
\end{enumerate}
\end{tcolorbox}
\vspace{0.1cm}
\caption{\textbf{Subspecialist Preference Evaluation Form.} Subspecialists were provided with two different responses (blinded) and asked to supply their preference (Response 1, Tie, or Response 2) for 9 different aspects of the response as well as the entire response as a whole. The same rubric was used for the direct comparison between AMIE's and the cardiologists' responses as well as between the cardiologists' responses with (assisted) and without (unassisted) access to AMIE's answers.}
\label{fig:Box3}

\begin{tcolorbox}[
    colback=black!5!white,
    colframe=black!60!white,
    title=\textbf{Subspecialist Individual Evaluation Form},
    fonttitle=\bfseries,
    arc=3mm,
    boxrule=1pt,
    bottomrule=2pt,
]
\footnotesize
\begin{enumerate}
  \item Are there any clinically significant errors?
  \item Does the answer contain any content it shouldn’t?
  \item Does the answer omit any content it shouldn’t?
  \item Does the answer contain any evidence of correct reasoning steps?
  \item Does the response contain any information that is inapplicable or inaccurate for any particular medical demographic?
\end{enumerate}
\end{tcolorbox}
\vspace{0.1cm}
\caption{\textbf{Subspecialist Individual Evaluation Form.} In addition to their direct side-by-side preferences (AMIE vs. Cardiologists and Assisted cardiologist vs. Unassisted cardiologist), subspecialists also independently answered each of these 5 questions (Yes/No) for AMIE's and the cardiologists' responses.}
\label{fig:Box2}
\end{figure}

\subsubsection{Statistical analysis}
\label{sec:stats}
For each preference or independent rating, we determined significance by using bootstrapping to estimate the sampling distribution of our test statistic~\citep{efron1992bootstrap}. For individual evaluation criteria, this test statistic was computed as (\% yes for AMIE - \% yes for Cardiologist), while for the preference ratings, it was computed as (\% AMIE preferred - \% Cardiologist preferred) or (\% Assisted preferred - \% Unassisted preferred). Note that we discarded ties for the purpose of this analysis. We created 10,000 re-samples of our data and calculated these statistic for each re-sample. We then determined significance by observing whether a difference of zero fell below the fifth percentile of the bootstrap distribution.

\section{Results}

204 consecutive patients were assessed by both general cardiologists and AMIE. The median age of patients was 59 years (range: 18-96 years). The number and percentage of patients with available clinical text data for each test was as follows: CMR: 121 (59.3\%), CPX: 115 (56.4\%), resting TTE: 172 (84.3\%), exercise TTE: 131 (64.2\%), ECG: 188 (92.2\%), ambulatory holter monitor: 151 (74.0\%), and genetic testing: 147 (72.0\%) (see Table~\ref{tab:clinicaltext}). Of the 204 patients, 75 (36.8\%) had a variant adjudicated to be pathogenic or likely pathogenic as per American College of Medical Genetics and Genomics (ACMG) interpretation of variants criteria~\citep{Richards2015-sn}. 

\begin{table}[h!]
\centering
\footnotesize
\resizebox{0.6\textwidth}{!}{
\begin{tabular}{cc}
\toprule
\textbf{} & \textbf{Number (\%)} \\ \hline
Mean age & 59.0 (range: 18-96) \\
Hypertrophic cardiomyopathy & 41 (20.1\%)\\
Left ventricular noncompaction & 40 (19.6\%)\\ 
Dilated cardiomyopathy & 16 (7.8\%) \\
Arrhythmogenic cardiomyopathy & 21 (10.3\%) \\
Ischemic cardiomyopathy & 22 (10.8\%) \\
Other genetic & 22 (10.8\%) \\
Non-genetic/general & 42 (20.6\%) \\
\hline
Cardiac MRI (CMR) & 121 (59.3\%) \\
Cardiopulmonary stress test (CPX) & 115 (56.4\%) \\ 
Resting transthoracic echocardiogram (TTE) & 172 (84.3\%)  \\ 
Exercise transthoracic echocardiogram (TTE) & 131 (64.2\%)  \\
Electrocardiogram (ECG) & 188 (92.2\%)  \\ 
Ambulatory Holter monitor & 151 (74.0\%) \\
Genetic testing & 147 (72.0\%) \\
\bottomrule
\end{tabular}}
\vspace{0.2cm}
\caption{\textbf{Clinical text data availability across patients.} The median age was 59 years (range: 18-96 years) across the 204 test cases. Clinical text data spans 7 different modalities: cardiac MRI, cardiopulmonary stress test, resting transthoracic echocardiogram, exercise transthoracic echocardiogram, electrocardiogram, ambulatory Holter monitor, and genetic testing.}
\label{tab:clinicaltext}
\end{table}

\clearpage
\subsection{Direct preference: AMIE vs. Cardiologist}
\begin{figure}[htbp!]
    \centering
    \includegraphics[width=\textwidth]{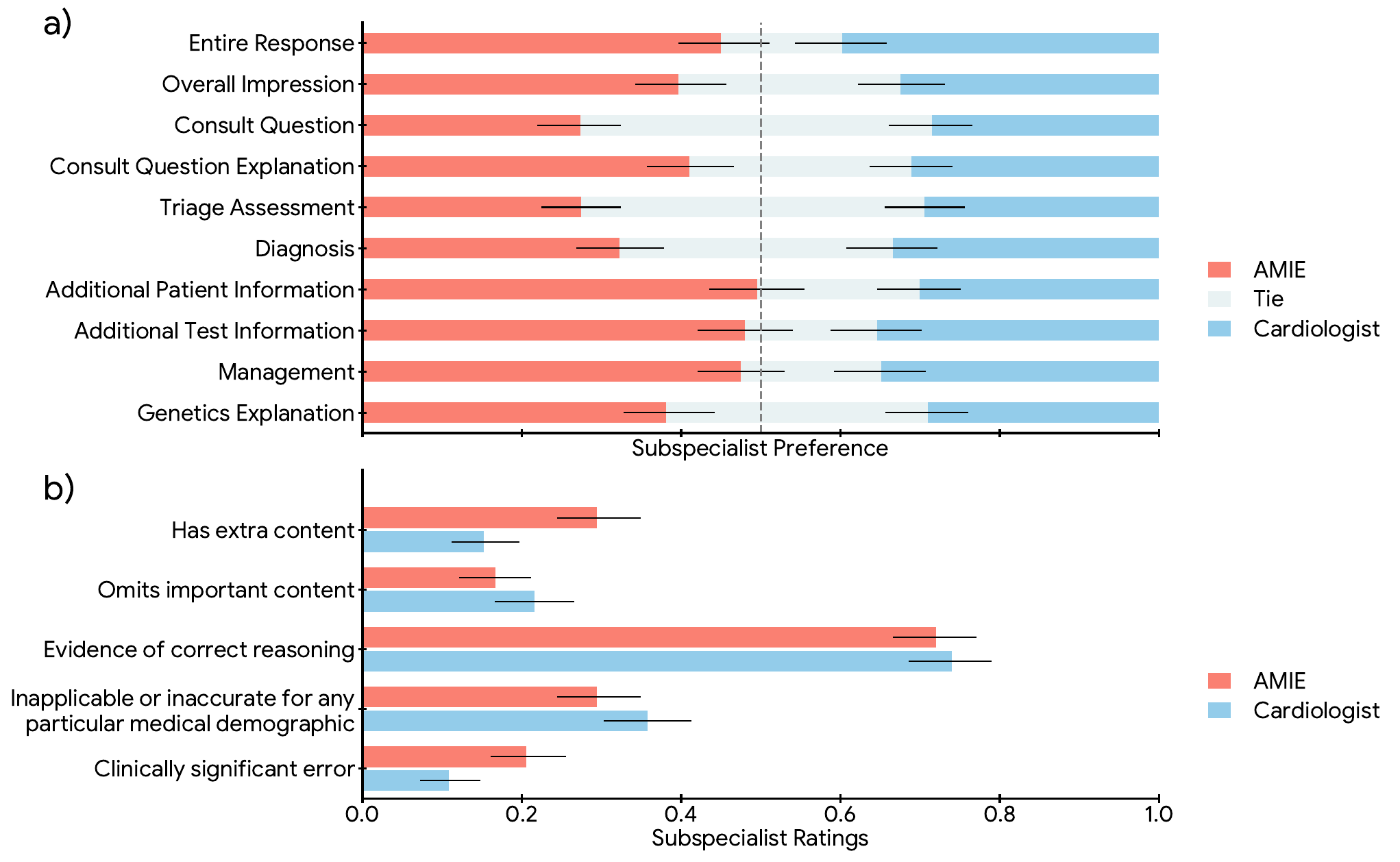}%
    \caption{\textbf{a) Preference between AMIE and cardiologist responses.} AMIE responses are preferred over the cardiologist responses for 5 of 10 domains (Consult Question Explanation, Additional Patient Information, Additional Test Information, Management, and Genetics Explanation) and non-inferior for the rest. \textbf{b) Individual assessment of AMIE and cardiologist responses.} Bars indicate the proportion of `yes' responses for each of the questions in \cref{fig:Box2}. AMIE's responses more often have extra content and clinically significant errors, while the cardiologists' responses more often are inapplicable for particular medical demographics.}%
    \label{fig:direct}%
\end{figure}

The subspecialist evaluators first directly compared AMIE's assessment to the general cardiologists' assessments using the evaluation form in \cref{fig:Box3}. For the 204 patients, across the 10 evaluation domain considered, AMIE was rated superior to general cardiologists across five domains while being equivalent for the remaining domains (Figure~\ref{fig:direct}a and Table~\ref{tab:direct}). The domains in which AMIE responses were preferred were: `consult question explanation', `additional patient information',  `additional test information', `management', and `genetic explanation'.

\subsection{Individual assessment}
The subspecialist evaluators evaluated AMIE and the general cardiologist responses individually on the set of five `yes'/`no' question in~\cref{fig:Box2}. As shown in~\cref{fig:direct}b and \cref{tab:individual}, AMIE's responses were less likely to omit relevant content, though this difference was not statistically significant. AMIE responses were more likely to contain unnecessary extra content or a clinically significant error. On the other hand, AMIE's responses were less likely to be inapplicable/inaccurate for particular medical demographics, and were rated equivalent to general cardiologist responses in evidence of correct reasoning.

\subsection{Direct preference: Cardiologist before and after seeing AMIE's response}
\begin{figure}[hbtp!]
    \centering
    \includegraphics[width=\textwidth]{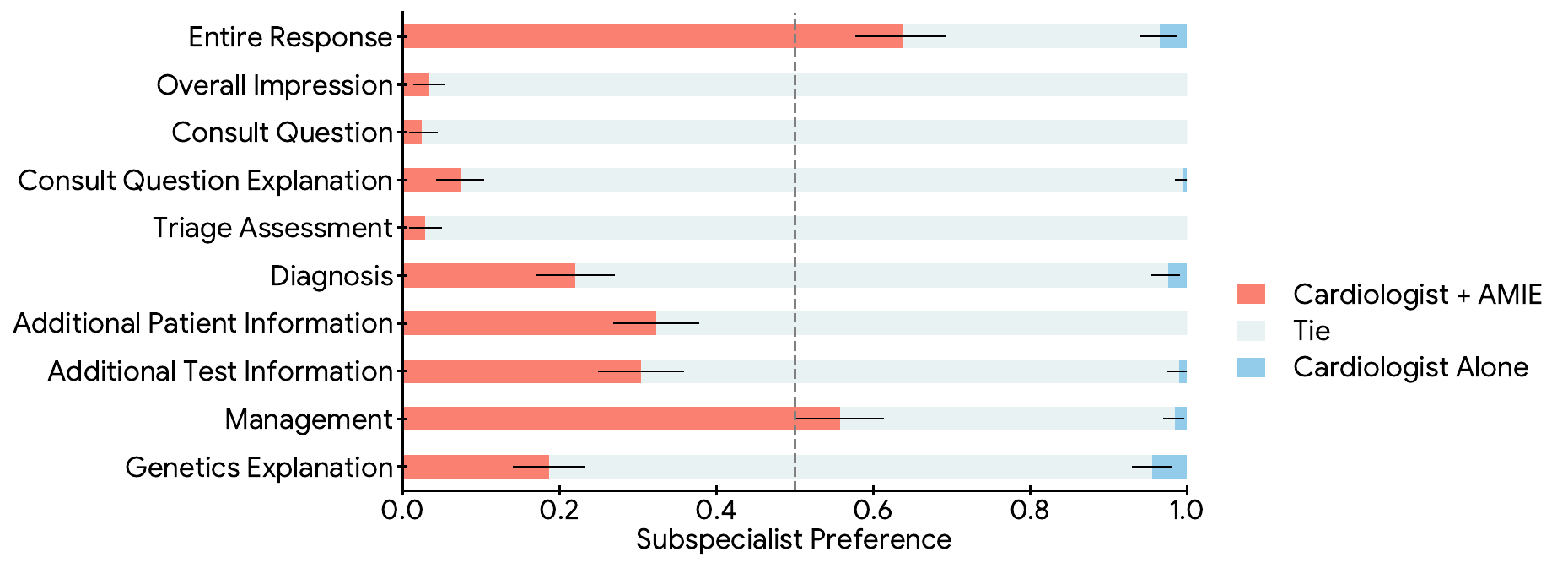}
    \caption{\textbf{Preference between cardiologist responses with and without access to AMIE's response.} For all 10 domains, the responses with access to AMIE's assessment were preferred over responses from the cardiologists alone.}%
    \label{fig:assist}%
\end{figure}

Of the 204 patient assessments, 195 of the assessments (95.6\%) were changed by the general cardiologists after seeing AMIE's response. While for a few assessments this consisted of relatively inconsequential wording changes, for many cases this revision substantially improved the appropriateness of the diagnosis, management plan, and/or genetic testing interpretation. When considering the entire response, the assisted `Cardiologist + AMIE' responses were ranked over cardiologists' responses without AMIE, preferred 63.7\% of the time, compared with 3.4\% of the time for responses by unassisted general cardiologists. Across the remaining 9 specific domains, the AMIE-assisted responses were preferred for all domains when directly compared to the general cardiologists alone, though `Tie' was the most common evaluation for 8 of the 10 domains (see Figure~\ref{fig:assist} and Table~\ref{tab:assist}).

The domains that had the greatest improvement were `management', `additional patient information', and `additional test information'. In particular, the management domain, which improved the most with AMIE assistance, had many cited causes for improvements including newer medications for rare diseases (e.g., cardiac myosin inhibitors for hypertrophic cardiomyopathy), education around family screening, and identification of further risk factors for sudden cardiac death. 

\subsection{Analysis of clinical feedback}
To understand the rationale behind the preferences and individual ratings provided by subspecialists, we analyzed the free-text comments left by subspecialists for AMIE's and the general cardiologists' responses. We combined all of the feedback received and summarize the reasons for preference for both AMIE and Cardiologists. While AMIE and cardiologists had similar overall preferences (see \cref{fig:direct}), the types of feedback they each received were quite different; AMIE was seen as thorough, and sensitive to a broad differential diagnosis, while the general cardiologists were often seen as specific, and concise, but risked anchoring on a certain diagnosis (see \cref{fig:preference_comments}). In this way, the approach of AMIE and general cardiologist might complement each other: AMIE is thorough and sensitive, general cardiologist concise and specific - analogous to combining a sensitive initial diagnostic test with a highly specific confirmatory test. We quantified this difference in specialist feedback of AMIE and general cardiologist responses: 92\% of general cardiologist's clinically significant errors described by subspecialists were omission errors and only 8\% of errors related to unnecessary extra care. These omission errors were a combination of: missing the correct diagnosis, overlooking imaging findings suggestive of disease or requiring further investigation (e.g., `biventricular systolic dysfunction'), or not recommending the appropriate diagnostic or follow up tests (e.g., no family screening, or no holter to investigations and assessment of sudden cardiac death risk). In contrast, only 35.5\% of AMIE's errors were omission, however 64.9\% were related to suggestions of potentially unnecessary care. The majority of these suggestions were additional or repeat non-invasive tests (such as holter monitors, echocardiograms and cardiac MRIs), with  9.1\% of described errors related to unnecessary suggestions of invasive tests/procedures (such as transesophageal echocardiogram and EP study). Both AMIE and general cardiologists' clinically significant errors are described in \cref{fig:error_comments}. We also performed a similar analysis on the 5 individual assessment criteria, finding that the subspecialists described very different and often complementary strengths and weaknesses of AMIE and  cardiologists for each criteria (see \cref{fig:llm_summaries}).

\begin{figure}[htp]
\begin{tcolorbox}[
    colback=black!5!white,
    colframe=black!60!white,
    title=\textbf{Summary of Subpecialist comments for AMIE vs. Cardiologist preferences},
    fonttitle=\bfseries,
    arc=3mm,
    boxrule=1pt,
    bottomrule=2pt,
]
\footnotesize
\textbf{Prompt to Gemini 1.5 Flash:} ``Instructions: The following is a list of feedback for a set of pairs of cardiologist assessments of patients with potential cardiovascular disease.
Specialists reviewed the assessments of the cardiologists and answered which of the cardiologists set of assessments they preferred. These are the set of specialists comments: \textcolor{blue}{\{feedback\}}
Describe the main feedback for specialists of AMIE vs. the general cardiologist. Indicate some reasons for preference of each group (both AMIE and general cardiologist).''\\

\tcbline
\vspace{0.1cm}
\textbf{Main Feedback:}
\begin{itemize}
\item AMIE assessments were often considered more thorough and comprehensive in their diagnostic workup and management plans. Specialists frequently praised AMIE for considering a broader differential diagnosis, exploring various potential etiologies (e.g., infiltrative, arrhythmogenic cardiomyopathies), and recommending a more detailed investigation.
\item General cardiologists were sometimes criticized for being too focused on a specific diagnosis too early. Specialists found that some general cardiologists jumped to conclusions, didn't consider all the available data, or overinterpreted certain findings. They also noted instances where general cardiologists missed crucial information or recommendations (e.g., anticoagulation).
\item Both AMIE and general cardiologists were sometimes found to miss certain diagnoses or misinterpret genetic findings. This highlights the inherent complexity of the field and the challenges in interpreting sometimes subtle or ambiguous clinical data.
\item General cardiologists were sometimes seen as being more concise, while AMIE was sometimes seen as being too detailed or jumping to conclusions. This suggests a balance needs to be struck between thoroughness and conciseness.
\end{itemize}

\textbf{Reasons for Preference of AMIE}
\begin{itemize}
\item Broader Differential Diagnosis: AMIE assessments often considered a wider range of potential diagnoses, leading to more comprehensive evaluations.
\item Thorough Workup: AMIE tended to recommend a more detailed diagnostic workup, including more specific tests and investigations.
\item Better Management Strategies: AMIE often proposed more comprehensive and appropriate management plans, including long-term surveillance and risk stratification.
\item Better Incorporation of Genetic Information: AMIE assessments were often better at integrating genetic findings into the overall clinical picture and management plan.
\item Less Likely to Anchor on a Single Diagnosis Too Early: AMIE assessments were more likely to consider multiple possibilities before settling on a specific diagnosis.
\end{itemize}

\textbf{Reasons for Preference of General Cardiologist}
\begin{itemize}
\item Conciseness: Some specialists preferred the more concise nature of the general cardiologist's assessments, finding them easier to understand and follow.
\item Focus on Key Issues: In some cases, the general cardiologist's more focused approach was seen as beneficial, particularly when the clinical picture was relatively straightforward.
\item Avoiding Unnecessary Testing: Some specialists appreciated that general cardiologists were more cautious about recommending excessive testing, potentially avoiding unnecessary burden on patients.
\end{itemize}

\textbf{In Summary: }
The feedback suggests that AMIE assessments generally provided a more thorough and comprehensive approach to evaluating patients with potential cardiovascular disease. However, this was at the expense of longer responses that sometimes contained unnecessary information or a significant clinical error. There were instances where AMIE assessments were considered too detailed. General cardiologists, while often more concise, were occasionally criticized for being too focused on a specific diagnosis or missing important details.
\end{tcolorbox}
\vspace{0.2cm}
\caption{\textbf{LLM-generated summary of subspecialist comments for preference rating between AMIE and the cardiologists.} Here, we aggregated all subspecialist comments for their preference choices and used Gemini 1.5 Flash \cite{reid2024gemini} to summarize the reasons for preference of AMIE and the cardiologists. Subspecialists left feedback on 159 out of 204 assessment pairs (77.9\%). This LLM-generated summary of feedback was reviewed and lightly edited by a subspecialist to validate its relevance and accuracy.}
\label{fig:preference_comments}
\end{figure}

\subsection{Qualitative clinical applications}
\begin{figure}[htp!]
    \centering
    \includegraphics[width=\textwidth]{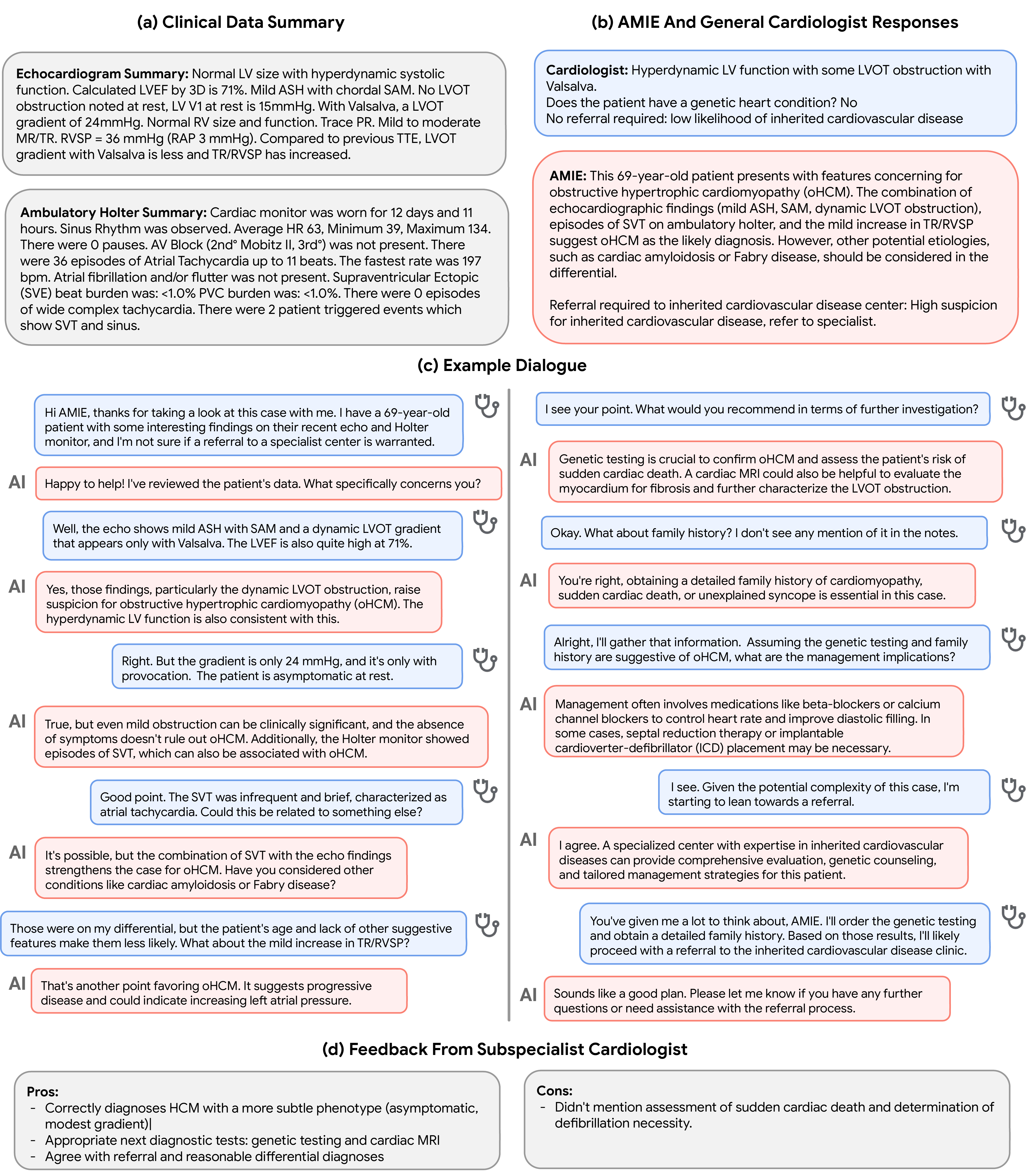}
    \vspace{0.2cm}
    \caption{\textbf{Dialogue between AMIE and a general cardiologist.} In this example, we prompted AMIE to generate how it would interact with a cardiologist. The patient has Hypertrophic Cardiomyopathy, but a somewhat subtle phenotype diagnosis, and divergent assessments from a general cardiologist and AMIE. This example shows how AMIE may assist a general cardiologist to manage this patient, highlighting that hypertrophic cardiomyopathy can present with a modest left ventricular outflow tract obstruction and can be asymptomatic. Therefore this patient should be referred to a specialist center.}%
    \label{fig:dialogue_pat19}%
\end{figure}

To explore potential future clinical uses of technology such as AMIE, we present four qualitative examples of how capabilities in dialogue could be utilized to communicate with patients or up-level generalists. The first hypothetical scenario in \cref{fig:dialogue_pat19} shows AMIE assisting a general cardiologist in the assessment of real-world clinical ECG and ambulatory Holter monitor text data (Figure~\ref{fig:dialogue_pat19}a). Figure~\ref{fig:dialogue_pat19}b shows the responses to this data by the general cardiologist and AMIE respectively, and Figure~\ref{fig:dialogue_pat19}c shows an example dialogue of AMIE assisting a general cardiologist. As Figure~\ref{fig:dialogue_pat19}b shows, the initial general cardiologist assessment of the patient had a low likelihood of genetic heart disease and no referral to a speciality genetic center was recommended. AMIE's response better reflected the subspecialist cardiologist assessment: that this patient's data suggests they may have hypertrophic cardiomyopathy and should be referred to a specialist center. This example shows how AMIE may assist a general cardiologist to manage this patient, highlighting that hypertrophic cardiomyopathy can present with a modest left ventricular outflow tract obstruction and can be asymptomatic. 

For the remaining three hypothetical scenarios, AMIE was given a prompt, akin to a ``one-line'' summary of a patient (\cref{tab:dialogue}) along with clinical data for the corresponding patient and asked to produce dialogues mirroring various potential use cases for AMIE: 
\begin{enumerate}
    \item AMIE reaches a diagnosis and then explains it to a patient (\cref{fig:dialogue_pat10});
    \item AMIE providing assistive dialogue for a general cardiologist (\cref{fig:dialogue_pat13});
    \item AMIE assumes the role of the specialist cardiologist and presents an assessment (\cref{fig:dialogue_pat2}).
\end{enumerate}

\section{Discussion}
In this study, we probe the ability of LLMs to provide additive support to generalists in the assessment of rare, life-threatening cardiac diseases that typically require subspeciality cardiac care. Further, we address the unmet need of evaluation of LLMs in specialist medical domains. To this end, we curate and open-source a de-identified real-world clinical dataset for patients suspected to have inherited cardiomyopathies and propose an evaluation rubric for the quality of diagnosis, triage and clinical management for such patients. Using this evaluation rubric, blinded subspecialists evaluate AMIE and general cardiologists before and after viewing AMIE's assessment. Across these 204 patients, AMIE is rated as equivalent (and for some domains, superior) in standalone performance, though with an increased rate of clinically significant errors. Furthermore, by improving assessment quality in over 60\% of patients, AMIE demonstrates significant potential to enhance general cardiologists' diagnoses and management of these challenging cases.

Our results demonstrate the feasibility of utilizing LLMs, specifically AMIE, an experimental research LLM optimized for diagnostic dialogue, to assess patients with rare and life-threatening cardiac conditions. Adapting AMIE to this subspecialist and rarified domain was highly data-efficient, leveraging iterative feedback from subspecialist experts to enhance the quality of AMIE’s responses using just 9 cases. This iterative process, combined with self-critique and the incorporation of search functionality, enabled AMIE to provide assessments equivalent (and for some domains, better) than those of general cardiologists. This contrasts with prior studies using generic, non-specialized LLMs, which did not achieve comparable clinical performance~\citep{hager2024evaluation}.

While AMIE provided more thorough assessments and exhibited fewer content omissions or biases, its increased comprehensiveness did come at the expense of a modest increase in clinically-significant errors. AMIE’s errors were generally additive, such as suggesting further investigations (i.e., such as regular monitoring with a cardiac MRI), whereas general cardiologist’s errors tended to be omission. Most commonly missing diagnoses, overlooking pathological imaging findings, and not recommending appropriate further investigations. In fact, 92\% of general cardiologist's errors were omission, whereas the majority of AMIE errors were related to suggestions of unnecessary care (65\%). Given the potential for AMIE's errors and the need for further prospective validation, our results do not support the deployment of LLMs like AMIE autonomously. The results instead indicate that a more appropriate use case for LLMs in this domain may be as assistive tools for generalist providers. AMIE is thorough and sensitive and its errors tend to be detectable and revolve around additional, potentially unnecessary testing. General cardiologists are specific, but prone to omission. In this way, AMIE's assistive value could be in thorough sensitive assessments, which then can be refined by cardiologists, who tend to be more specific. Importantly, AMIE's errors are detectable by cardiologists and few (9\%) related to invasive testing.

It has been shown that AI tools make different types of errors to clinicians in many areas of medicine, creating the potential for AI to have positive assistive impacts but also, counter-intuitively, creating the risk to worsen clinicians' performance~\citep{dvijotham2023enhancing}. This can occur through many mediating phenomena, including inappropriate over-reliance, under-reliance, automation bias, and inadequate onboarding~\citep{gaube2021ai, cai2019hello}. We therefore explored the effect on general cardiologists' assessments if they were allowed access to the AMIE responses, and found that presenting general cardiologists with AMIE's response improved their response over 60\% of the time, while decreasing quality for under 5\% of patients. This improvement was seen across all 10 evaluated domains, with management, genetic interpretation and advanced diagnostics the areas of most improvement. This finding suggests that general cardiologists were able to extract helpful and valid additive information from AMIE, while disregarding AMIE's errors. If further prospective research validates our findings, LLMs may have the potential to assist generalist in providing subspecialist care. The included dialogue scenarios qualitatively illustrate some of these potential use cases, from aiding generalists in their investigation and care to explaining complex information and delivering assessments to patients.

While further research could extrapolate our approach to a broader group of specialities, cardiology is a useful indicative example because it features a) highly preventable morbidity and mortality, b) a reliance on an array of clinical investigations (e.g., echocardiograms, electrocardiograms, ambulatory holter monitors, cardiac MRIs, cardiopulmonary stress tests, and, when appropriate, genetic testing), and c) a substantial deficit in cardiology workforce. Our findings are of particular note as access to subspecialist care is a global challenge. The American College of Cardiology has identified a `cardiology workforce crisis', with lack of access to subspeciality cardiologists an acute concern~\citep{Maddox2024-ci}. In the US, despite five HCM centers of excellence in both California and New York, there are none across 27 states~\citep{hcmcenter}. This has led to more than 60\% of HCM patients in the US undiagnosed, with estimates higher globally~\citep{Massera2023-og}. The propensity of inherited cardiomyopathy to cause sudden cardiac death (the leading cause of SCD in young adults~\citep{Ommen2024-em}), exacerbates the problem. Lack of access to appropriate care and long wait times can lead to preventable, premature mortality. LLMs may help identify undiagnosed cases, assist with triage and prioritization of urgent cases, and streamline management. In this way, LLMs could improve access to specific care, by assisting generalists.

Our study addresses a meaningful wider gap in prior literature. Prior research has evaluated LLMs in a number of different settings in medicine, from assessing quality in question-answering (spanning medical license examinations as well as open-ended medical questions) to clinical image and complex diagnostic challenges~\citep{singhal2023large, eriksen2023use, saab2024capabilities, nori2023can, singhal2023towards, yang2024pediatricsgpt,jin2024hidden, luk2024performance, katz2024gpt}. Other studies have investigated the utility and potential of LLMs in real-world clinical tasks such as clinical letter generation~\citep{ali2023using}, medical information communication~\citep{cox2023utilizing}, and medical summarization~\citep{patel2023chatgpt, van2024adapted}, and triaging mammograms and chest x-rays for tuberculosis~\citep{dvijotham2023enhancing}. Existing research on the performance of LLMs in medical subspecialties, such as cardiology~\citep{giannos2023evaluating}, ophthalmology~\citep{tao2024chatgpt}, gastroenterology~\citep{suchman2023chat}, neurology~\cite{giannos2023evaluating}, and surgery~\citep{beaulieu2024evaluating}, are also mostly limited to medical question-answering or examination benchmarking. A recent study~\cite{chen2023use} evaluated ChatGPT on providing accurate cancer treatment recommendations concordant with authoritative guidelines with fixed question prompts. Another study investigated the diagnostic and triage accuracy of the GPT-3 relative to physicians and laypeople using synthetic case vignettes of both common and severe conditions~\citep{levine2024diagnostic}. A 2024 study~\citep{jo2024assessing} compared GPT-4 performance with human experts on answering cardiology-specific questions from general users' web queries. Our study is not only one of the first scientific assessments of LLMs in subspecialty domains~\citep{kim2024assessing}, it is also, to our knowledge, one of the first to use real-world data, and make this data for LLM evaluation available open-source.

The existing literature shows mixed results for LLMs on clinical cases~\citep{van2024adapted, hager2024evaluation}. A recent study showed the potential and safety concerns of using LLM to provide on-demand consultation service that assists clinicians to provide bedside decision-making based on patient EHR data~\citep{callahan2021using}. A 2024 study assessed the ability of LLMs to diagnose abdominal pathologies and showed that LLMs were inferior to clinicians~\citep{hager2024evaluation}. The authors note that their results may be improved with fine-tuned LLMs. Though we do not fine-tune for this particular downstream task, our approach, which included using a medically-specialized LLM equipped with web search and a multi-step reasoning chain at inference time may explain our contrasting results. 

Currently LLMs are being utilised in many US health systems via their implementation in electronic medical records software~\cite{turner2023epic}. This implementation has occurred without similar scale of scientific evaluation: the benefits and the possible harms are only partially known~\citep{schoonbeek4835935completeness}. Our results serve as one step closer towards demonstrating the real-world utility of LLMs for subspecialty care.

Our study contains a number of important limitations and the findings should be interpreted with appropriate caution and humility. First, clinicians and LLMs were constrained to reviewing text-based reports of investigations, rather than the multimodal investigations themselves. This presents the possibility of upstream errors, but also limits the potential impact of such technology because a wide array of specialist expertise is still required to report each investigation in turn. Further, history and physical examination are indispensable components of real clinical practice, but they were not included in this study. This is a limitation of the applicability of our work and future studies should consider the settings in which there is prospective interaction with these patients. However, our approach does mirror an increasingly common way care is provided: an electronic consultant (E-consult)~\citep{Vimalananda2020-aq}. E-consults are similar to our results where text-only information is shared with a specialist for their assessment. Further limitations of our work include a biased sample of patients - patients were selected from one US center, using only English text. It is unclear how well our results will extrapolate to other non-US settings. Further, our dataset contained patients that were indeed referred for a suspicion of an inherited cardiac disease (correctly or incorrectly). A less biased population may be from a general cardiology clinic, where the prevalence of inherited disease is lower and with it possibly a higher chance of false positive referral rate. A similar limitation is our patients already had a number of cardiac diagnostic tests completed. To help identify undiagnosed cases, LLMs would have to be studied in populations with less complete cardiac investigations. There was insufficient demographic or regional variation in the single-centre population in our study to assess the potential for bias or health inequity, which is an important topic for AI systems in healthcare. This is crucial as prospective studies are considered as disparities are rife in the care of patients with inherited cardiomyopathies \citep{Ntusi2021-cl}. The possible assistive impacts we observed in this study were obtained only by presenting fixed AMIE outputs to clinicians and allowing them to edit a prior plan. Often the changes made by general cardiologists after AMIE assistance were minimal, but still clinically significant, such as adding family screening. Further research is required to better characterize and optimise the assistive user experience, and study the impact of dialogue between AMIE and the general cardiologist on the final composite outcome of clinician-AI teamwork. 

Similarly, our research did not explore the potential benefits and risks from the perspective of patients. The early potential here demonstrates an opportunity for participatory research including the patient perspective on many potentially different workflows that could be enabled for subspecialist consultation. Finally, our approach included only retrospective data from historic clinical cases. Prior to assessing safety for clinical implementation, prospective research is necessary.

While AMIE's performance was promising, there were notable areas for improvement that were highlighted by our evaluation rubric, including the presence of significant clinical errors, among longer responses than those obtained by general cardiologists. The complementary and assistive utility of the technology requires significant further study before it could be considered safe for real-world use, and there are many other considerations beyond the scope of this work including regulatory and equity research, and validation in a wider range of clinical environments.

In conclusion, AMIE, a research LLM-based AI system optimised for clinical and diagnostic dialogue, showed equivalence with general cardiologists in the assessment of patients with rare, life-threatening inherited cardiomyopathies. AMIE had was seen as less demographically biased, showed equivalent clinical reasoning, and was more thorough than general cardiologists, though this was at the expense of more errors. The most encouraging result was the potential for LLMs to be a useful clinical aid; when general cardiologists leveraged AMIE to refine their own responses, all domains of their assisted responses were preferred over their unassisted responses. Furthermore, these changes resulted in a meaningful, positive improvement for the entire response in over 60\% of patients.

\section{Ethics approval}
The clinical subspecialist evaluator component of this research involved the participation of physicians. This study adhered to the principles outlined in the Declaration of Helsinki. Informed consent was obtained from each physician before their participation. This study used only retrospective, de-identified data that fell outside the scope of institutional review board oversight.

\section{Reporting summary}
Further information on research design is available in the separate reporting summary

\subsubsection*{Acknowledgments}
We would like to thank John Lugo for his instrumental support. We would also like thank Jonathan Gortat and Scott Edmiston for their support, particularly in facilitating the open-sourcing of our data. We thank Fan Zhang and Cían Hughes for their comprehensive review and detailed feedback on the manuscript.

\subsubsection*{Data availability}
 Data consists of clinical test text data (ECGs, CMRs, rest and stress TTEs, ambulatory holter monitors, cardiopulmonary stress tests). All data is available open-sourced, available at \url{https://redivis.com/datasets/1z3x-2354972da?v=next}. Data is licensed under open-source license CC 4.0.  

\subsubsection*{Code availability} AMIE is an LLM based research AI system for diagnostic dialogue. We are not open-sourcing model code and weights due to the safety implications of unmonitored use of such a system in medical settings. In the interest of responsible innovation, we will be working with research partners, regulators, and providers to validate and explore safe onward uses of AMIE. For reproducibility, we have documented technical deep learning methods in~\citep{tu2024towards} while keeping the paper accessible to a clinical and general scientific audience.  Our work builds upon PaLM 2, for which technical details have been described extensively in the technical report~\citep{google2023palm2}.

\subsubsection*{Competing interests}
This study was funded by Alphabet Inc and/or a subsidiary thereof (`Alphabet'). Anil Palepu, Khaled Saab, Wei-Hung Weng, Yong Cheng, Philip Mansfield, Dale Webster, Juraj Gottweis, Joelle Barral, Ryutaro Tanno, Mike Schaekermann, S. Sara Mahdavi, Vivek Natarajan, Alan Karthikesalingam, and Tao Tu are employees of Alphabet and may own stock as part of the standard compensation package. D.S.K. reports grant support from Amgen. D.S.K. is supported by the Wu-Tsai Human Performance Alliance as a Clinician-Scientist Fellow, the Stanford Center for Digital Health as a Digital Health Scholar, the Robert A. Winn Diversity in Clinical Trials Career Development Award, and NIH 1L30HL170306. EA reports advisory board fees from Apple and Foresite Labs. EA has ownership interest in SVEXA, Nuevocor, DeepCell, and Personalis, outside the submitted work. EA is a board member of AstraZeneca. JOS is supported by the Wu-Tsai Human Performance Alliance as a Clinician-Scientist Fellow and has had consultancy relationships with Google AI (outside the current work). VP has consulting and advisory relationships with BioMarin, Lexeo Therapeutics and Viz.ai and receives funding from BioMarin, the John Taylor Babbitt Foundation, the Sarnoff Cardiovascular Research Foundation, and NHLBI R01HL168059 and K08HL143185. SSJ reports consulting fees from Bristol Myers Squibb, ARTIS Ventures, and Broadview Ventures outside of the submitted work. All other authors declare no conflicts of interests. 

\newpage
\setlength\bibitemsep{3pt}
\printbibliography
\balance
\clearpage

\end{refsection}

\newpage
\begin{refsection}

\clearpage

\renewcommand{\thesection}{A.\arabic{section}}
\renewcommand{\thefigure}{A.\arabic{figure}}
\renewcommand{\thetable}{A.\arabic{table}} 
\renewcommand{\theequation}{A.\arabic{equation}} 
\renewcommand{\theHsection}{A\arabic{section}}

\setcounter{section}{0}
\setcounter{figure}{0}
\setcounter{table}{0}
\setcounter{equation}{0}

\noindent \textbf{\LARGE{Appendix}}\\
\normalfont

In the following sections, we provide:
\begin{itemize}[leftmargin=1.5em,rightmargin=0em]
\item An example case response from AMIE (\cref{sec:example_response}).
\item Additional information of subspecialists' evaluation on AMIE and the general cardiologist's responses (\cref{sec:rating_tables}).
\item LLM-generated summaries of the comments left by subspecialists for the AMIE and cardiologist individual assessments (\cref{sec:llm_summary_of_feedback}).
\item LLM-generated comparison of the types of clinical errors made by AMIE and the general cardiologists
\item Three scenarios and dialogue examples to illustrate potential applications of medical LLMs like AMIE in this domain (\cref{sec:additional_dialogues}).

\end{itemize}


\newpage

\section{Example model response}
\label{sec:example_response}

\begin{figure}[htp!]
    \centering
    \includegraphics[width=\textwidth]{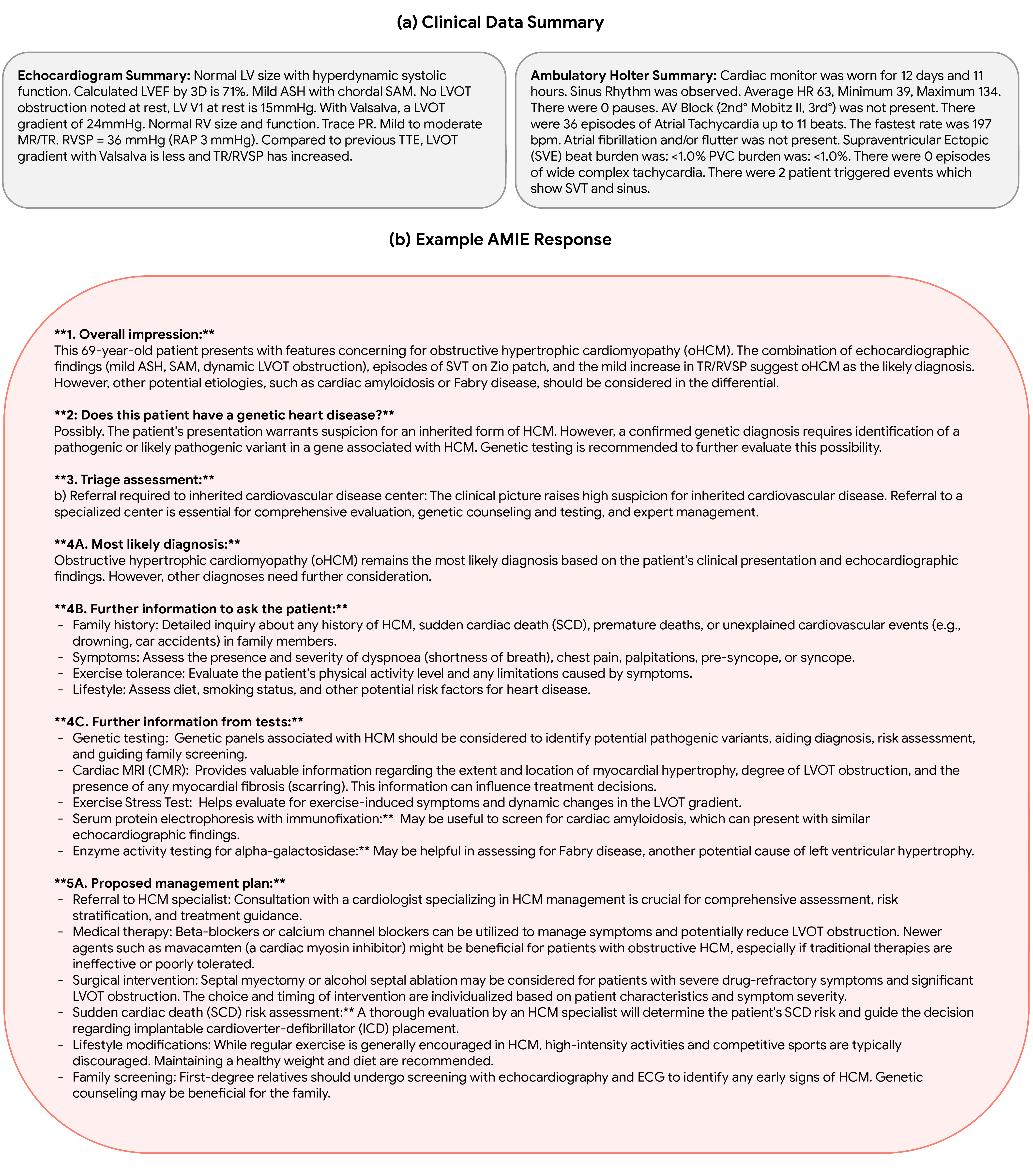}
    \vspace{0.2cm}
    \caption{\textbf{Example model response.} \textbf{a)} Summaries of the clinical data provided to AMIE. \textbf{b)} The response provided by AMIE to the questions in \cref{fig:Assessment}.}
    \label{fig:example_response}
\end{figure}

\newpage

\section{Additional evaluation information}
\label{sec:rating_tables}

Here we present detailed results from subspecialist evaluators including: the preference between AMIE and cardiologist responses, the individual assessment of AMIE and cardiologist responses, and the preference between cardiologist responses with and without access to AMIE’s response.

{\renewcommand{\arraystretch}{1.35}
\begin{table}[htbp!]
\footnotesize
\centering
\resizebox{\textwidth}{!}{%
\begin{tabular}{p{3cm}cccc}
\toprule
\textbf{Domain} & \textbf{AMIE \%} & \textbf{Cardiologist \%}& \textbf{Tie \%}& \textbf{AMIE - Cardiologist}\\
& (90\% CI) & (90\% CI) & (90\% CI) & (90\% CI)\\ 
\hline
Entire Response & \textbf{45.1} (39.2, 51.0) & 39.7 (33.8, 45.1) & 15.2 (11.3, 19.6) & 5.4 (-4.9, 16.2)\\ \hline
Overall Impression & \textbf{39.7} (34.3, 45.6) & 32.4 (27.0, 37.7) & 27.9 (23.0, 33.3) & 7.4 (-2.5, 17.2)\\ \hline
Consult Question & 27.5 (22.5, 32.8) & 28.4 (23.5, 33.8) & \textbf{44.1} (38.2, 50.0) & -1.0 (-9.8, 7.4)\\ \hline
Consult Question Explanation & \textbf{41.2} (35.3, 47.1) & 30.9 (25.5, 36.3) & 27.9 (23.0, 33.3) & \textbf{10.3} (0.5, 20.1)\\ \hline
Triage Assessment & 27.5 (22.5, 32.8) & 29.4 (24.5, 34.8) & \textbf{43.1} (37.7, 48.6) & -1.9 (-10.3, 6.9)\\ \hline
Diagnosis & 32.3 (27.0, 37.7) & 33.3 (27.9, 38.7) & \textbf{34.3} (28.9, 39.7) & -1.0 (-10.3, 8.3)\\ \hline
Additional Patient Information & \textbf{49.5} (43.6, 55.4) & 29.9 (24.5, 35.3) & 20.6 (16.2, 25.5) & \textbf{19.7} (9.8, 29.9)\\ \hline
Additional Test Information & \textbf{48.0} (42.2, 53.9) & 35.2 (29.9, 40.7) & 16.7 (12.7, 21.1) & \textbf{12.8} (2.5, 23.0)\\ \hline
Management & \textbf{47.5} (42.2, 53.4) & 34.7 (29.4, 40.2) & 17.7 (13.2, 22.1) & \textbf{12.8} (2.5, 23.0)\\ \hline
Genetics Explanation & \textbf{38.2} (32.8, 43.6) & 28.9 (23.5, 34.3) & 30.4 (25.5, 35.8) & \textbf{9.3} (0.0, 18.6)\\ 
\bottomrule
\end{tabular}}
\vspace{0.2cm}
\caption{\textbf{Preference rating between cardiologist and AMIE responses.} For each domain, we computed the proportion of the preferred responses and the most selected choice (`AMIE', `Cardiologist', or `Tie') as well as significant `AMIE' - `Cardiologist' differences are bolded. The difference between the proportion of AMIE and cardiologist preferences is shown on the right. For 5 of the 10 domains, AMIE's response was preferred over the response from cardiologists ($\alpha < 0.05$); AMIE was equivalent in the other 5 domains.}
\label{tab:direct}
\end{table}}

{\renewcommand{\arraystretch}{1.2}
\begin{table}[htbp!]
\footnotesize
\centering
\resizebox{\textwidth}{!}{%
\begin{tabular}{p{6cm}ccc}
\toprule
\textbf{Domain} & \textbf{AMIE \%} & \textbf{Cardiologist \%} & \textbf{AMIE - Cardiologist}\\
& (90\% CI) & (90\% CI) & (90\% CI)\\
\hline
Has extra content & \textbf{29.4} (24.5, 34.8) & 15.2 (11.3, 19.6) & \textbf{14.2} (7.8, 20.6)\\ \hline
Omits important content & 16.7 (12.3, 21.1) & \textbf{21.6} (16.7, 26.5) & -4.9 (-11.3, 1.5)\\ \hline
Evidence of correct reasoning & 72.0 (66.7, 77.0) & \textbf{74.0} (68.6, 78.9) & -2.0 (-7.4, 3.4)\\ \hline
Inapplicable or inaccurate for any\newline particular medical demographic & 29.4 (24.5, 34.8) & \textbf{35.8} (30.4, 41.2) & \textbf{-6.4} (-10.8, -2.0)\\ \hline
Clinically significant error & \textbf{20.6} (16.2, 25.5) & 10.8 (7.4, 14.7) & \textbf{9.8} (4.4, 15.2)\\ 
\bottomrule
\end{tabular}}
\vspace{0.2cm}
\caption{\textbf{Individual assessment of cardiologist and AMIE responses.} The center columns indicate proportion of `Yes' responses for each question in \cref{fig:Box2} (with higher proportions and significant differences bolded) for AMIE and the cardiologists, respectively, while the `AMIE - Cardiologist' column indicates this difference (with significant differences bolded). AMIE's responses more often have extra content and clinically significant errors, while the cardiologists' responses more often are inapplicable for particular medical demographics.}
\label{tab:individual}
\end{table}}

{\renewcommand{\arraystretch}{1.35}
\begin{table}[htbp!]
\footnotesize
\centering
\resizebox{\textwidth}{!}{%
\begin{tabular}{p{3cm}cccc}
\toprule
\textbf{Domain} & \textbf{Unassisted \%} & \textbf{Assisted \%} & \textbf{Tie \%} & \textbf{Assisted - Unassisted}\\ & (90\% CI) & (90\% CI) & (90\% CI) & (90\% CI)\\
\hline 
Entire Response & 3.4 (1.5, 5.9) & \textbf{63.7} (57.8, 69.1) & 32.8 (27.5, 38.2) & \textbf{60.3} (53.4, 66.7)\\ \hline
Overall Impression & 0.0 (0.0, 0.0) & 3.4 (1.5, 5.4) & \textbf{96.6} (94.6, 98.5) & \textbf{3.4} (1.5, 5.4) \\ \hline
Consult Question & 0.0 (0.0, 0.0) & 2.5 (1.0, 4.4) & \textbf{97.5} (95.6, 99.0) & \textbf{2.5} (1.0, 4.4) \\ \hline
Consult Question Explanation & 0.5 (0.0, 1.5) & 7.4 (4.4, 10.3) & \textbf{92.2} (88.7, 95.1) & \textbf{6.9} (3.9, 10.3) \\ \hline
Triage Assessment & 0.0 (0.0, 0.0) & 2.9 (1.0, 4.9) & \textbf{97.1} (95.1, 99.0) & \textbf{2.9} (1.0, 4.9) \\ \hline
Diagnosis & 2.4 (1.0, 4.4) & 22.0 (17.2, 27.0) & \textbf{75.5} (70.6, 80.4) & \textbf{19.6} (14.2, 25.0)\\ \hline
Additional Patient Information & 0.0 (0.0, 0.0) & 32.3 (27.0, 37.7) & \textbf{67.7} (62.3, 73.0) & \textbf{32.3} (27.0, 37.7)\\ \hline
Additional Test Information & 1.0 (0.0, 2.5) & 30.4 (25.0, 35.8) & \textbf{68.6} (63.2, 74.0) & \textbf{29.4} (24.0, 34.8) \\ \hline
Management & 1.5 (0.5, 2.9) & \textbf{55.8} (50.0, 61.3) & 42.7 (37.3, 48.5) & \textbf{54.3} (48.0, 60.3) \\ \hline
Genetics Explanation & 4.4 (2.0, 6.9) & 18.6 (14.2, 23.0) & \textbf{77.0} (72.1, 81.9) & \textbf{14.2} (8.8, 19.6) \\
\bottomrule
\end{tabular}}
\vspace{0.2cm}
\caption{\textbf{Preference between cardiologist responses with and without access to AMIE's response.} For each domain, the most selected choice (`Unassisted', `Tie', or `Assisted') as well as significant `Assisted' - `Unassisted' differences are bolded. For all 10 domains, the response with access to AMIE's assessment (`Assisted') was preferred over the response from cardiologists alone (`Unassisted'). Furthermore, for 2 of the domains, `Management' and `Entire Response', `Assisted' was also chosen more than `Tie'. Note that the subspecialist evaluators indicated that over 63\% of cardiologist responses improved after seeing AMIE's response, while under 4\% became worse.}
\label{tab:assist}
\end{table}}

\clearpage
\section{Summary of subspecialist free-text comments for individual assessments}
\label{sec:llm_summary_of_feedback}

\begin{figure}[htp!]
    \centering
    \includegraphics[width=\textwidth]{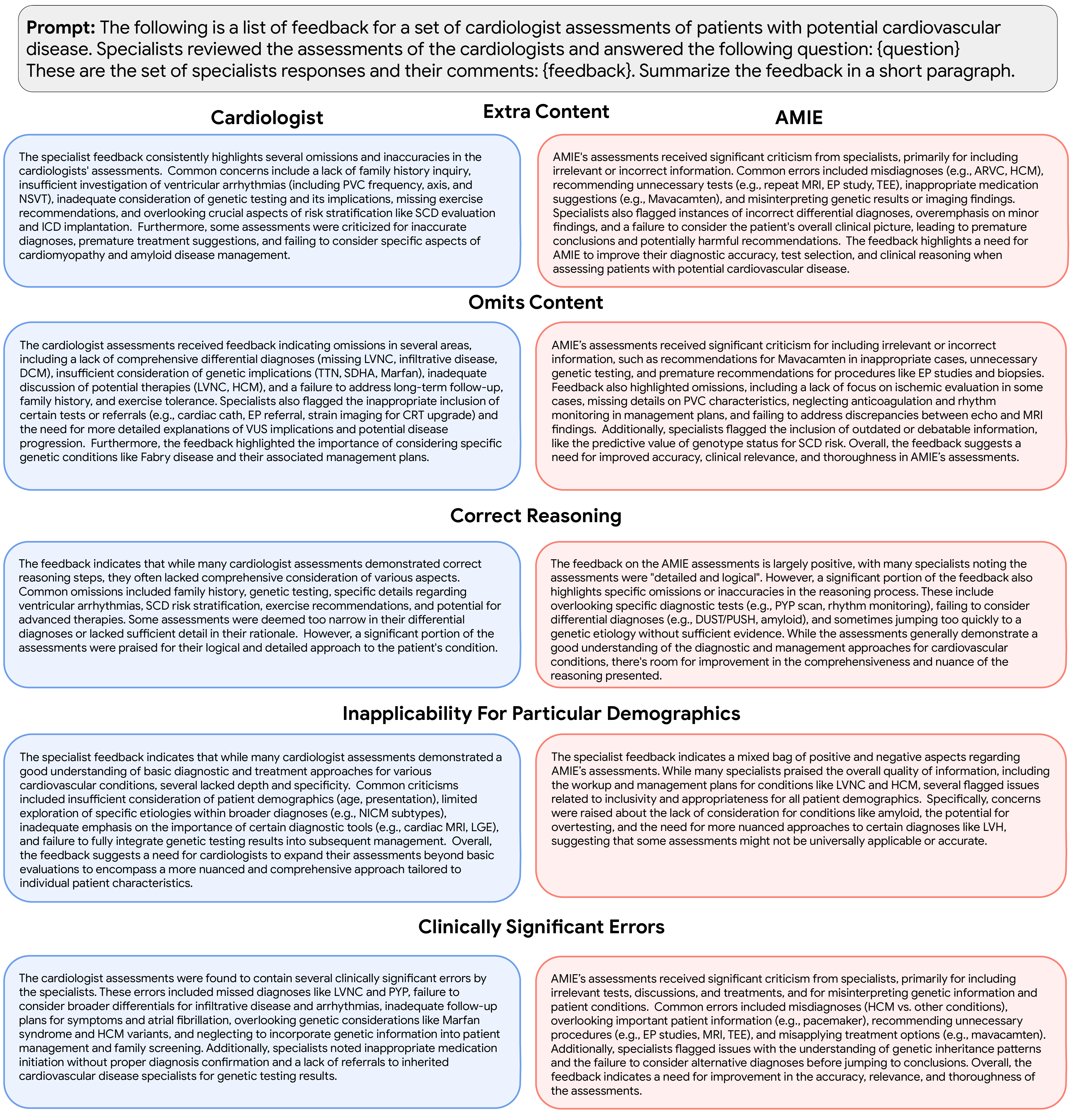}
    \vspace{0.2cm}
    \caption{\textbf{LLM-generated summaries of subspecialist comments to AMIE and cardiologist assessments.} We took all free-text comments from subspecialists for each of the 5 questions in the individual assessment (\cref{fig:Box2}) and asked Gemini 1.5 Flash to summarize the feedback for the cardiologists (left, blue boxes) and for AMIE (right, red boxes). Note that not every assessment received a comment from the subspecialists, and there is likely bias in which assessments they chose to comment on. Proportion of general cardiologist responses with comments: Extra Content: 15.7\%, Omits Content: 20.6\%, Correct Reasoning: 36.8\%, Inapplicability for particular demographics: 9.3\%, Clinically Significant Errors: 7.4\%. Proportion of AMIE responses with comments: Extra Content: 29.4\%, Omits Content: 16.2\%, Correct Reasoning: 29.4\%, Inapplicability for particular demographics: 5.9\%, Clinically Significant Errors: 18.1\%.}
    \label{fig:llm_summaries}
\end{figure}

\clearpage
\section{Summary of Clinically Significant Errors}

\begin{figure}[htp!]
\begin{tcolorbox}[
    colback=black!5!white,
    colframe=black!60!white,
    title=\textbf{Summary of AMIE's and Cardiologists' Clinically Significant Errors},
    fonttitle=\bfseries,
    arc=3mm,
    boxrule=1pt,
    bottomrule=2pt,
]
\footnotesize
\textbf{Prompt to Gemini 1.5 Flash:} ``Instructions: The following is a list of feedback regarding the clinically significant errors for a set of cardiologist assessments of patients with potential cardiovascular disease.
Specialists reviewed the assessments of the cardiologists and answered which of the cardiologists set of assessments they preferred.
These are the set of specialists comments for AMIE: \textcolor{blue}{\{amie feedback\}}\textcolor{black}. These are the set of specialists comments for the general cardiologist: \textcolor{blue}{\{card feedback\}}\textcolor{black}.  Describe the main feedback for specialists of AMIE vs the general cardiologist. Indicate key themes of clinically significant errors for each group (both AMIE and general cardiologist).''

\tcbline
\textbf{AMIE: Key Themes of Clinically Significant Errors:}
\begin{itemize}
\item Over-reliance on advanced testing and interventions: AMIE frequently recommended unnecessary tests like MRI, PET, TEE, and invasive procedures like biopsies and catheterizations, even when the clinical picture didn't warrant them. It also prematurely suggested treatments like mavacamten, even before a definitive diagnosis was established.
\item Misinterpretation of genetic information: AMIE often misinterpreted genetic findings, such as mistaking carrier status for disease, incorrectly interpreting VUS results, and failing to understand the implications of negative genetic tests in the context of a strong clinical phenotype.
\item Inaccurate diagnoses and management plans: AMIE made incorrect diagnoses, particularly in differentiating between different types of cardiomyopathy. It also formulated inappropriate management plans, such as recommending beta-blockers in ARVC or missing the need for anticoagulation in certain cases.
\item Lack of understanding of disease progression and natural history: AMIE demonstrated a lack of understanding of the natural history of some conditions, leading to unnecessary interventions or overlooking important aspects of long-term management. For example, it recommended repeat Holter monitoring when one had already been performed or discussed pregnancy in a 50-year-old patient.
\end{itemize}

\textbf{General Cardiologist: Key Themes of Clinically Significant Errors}
\begin{itemize}
\item Failure to consider a broad differential diagnosis: The general cardiologist often failed to consider a wide range of potential diagnoses, particularly missing rarer conditions like LVNC or infiltrative cardiomyopathies.
\item Incomplete workup: The general cardiologist often omitted crucial investigations like cardiac MRI, which would have been helpful in establishing a diagnosis.
\item Inadequate follow-up and genetic counseling: The cardiologist often failed to recommend appropriate long-term follow-up, including monitoring for disease progression and managing potential complications like atrial fibrillation. It also lacked in recommending genetic counseling and family screening when genetic variants were identified.
\item Lack of integration of genetic information into management: Even when genetic testing revealed pathogenic variants, the general cardiologist often failed to incorporate this information into the management plan, such as failing to refer to specialists or initiate appropriate cascade screening for family members.
\end{itemize}

\textbf{In Summary:}
AMIE's errors tended to be more related to over-testing, over-treatment, and misinterpretation of complex data, potentially reflecting a tendency to prioritize advanced technology over clinical judgment. The general cardiologist's errors were more focused on incomplete workup, limited differential diagnosis, and inadequate follow-up, possibly suggesting a more conservative approach and a lack of familiarity with rarer conditions and the latest guidelines for genetic testing and management.

\end{tcolorbox}
\vspace{0.2cm}
\caption{\textbf{LLM-generated summary of AMIE and the cardiologists clinically significant errors.} Here, we aggregated subspecialist descriptions of the clinically significant errors, using Gemini 1.5 Flash \cite{reid2024gemini} to summarize and compare the types of errors made by AMIE and the General Cardiologists. This LLM-generated summary was reviewed by a subspecialist to validate its relevance and accuracy.}
\label{fig:error_comments}
\end{figure}

\clearpage
\section{Additional dialogue examples}
\label{sec:additional_dialogues}

\begin{figure}[htp!]
    \centering
    \includegraphics[width=\textwidth]{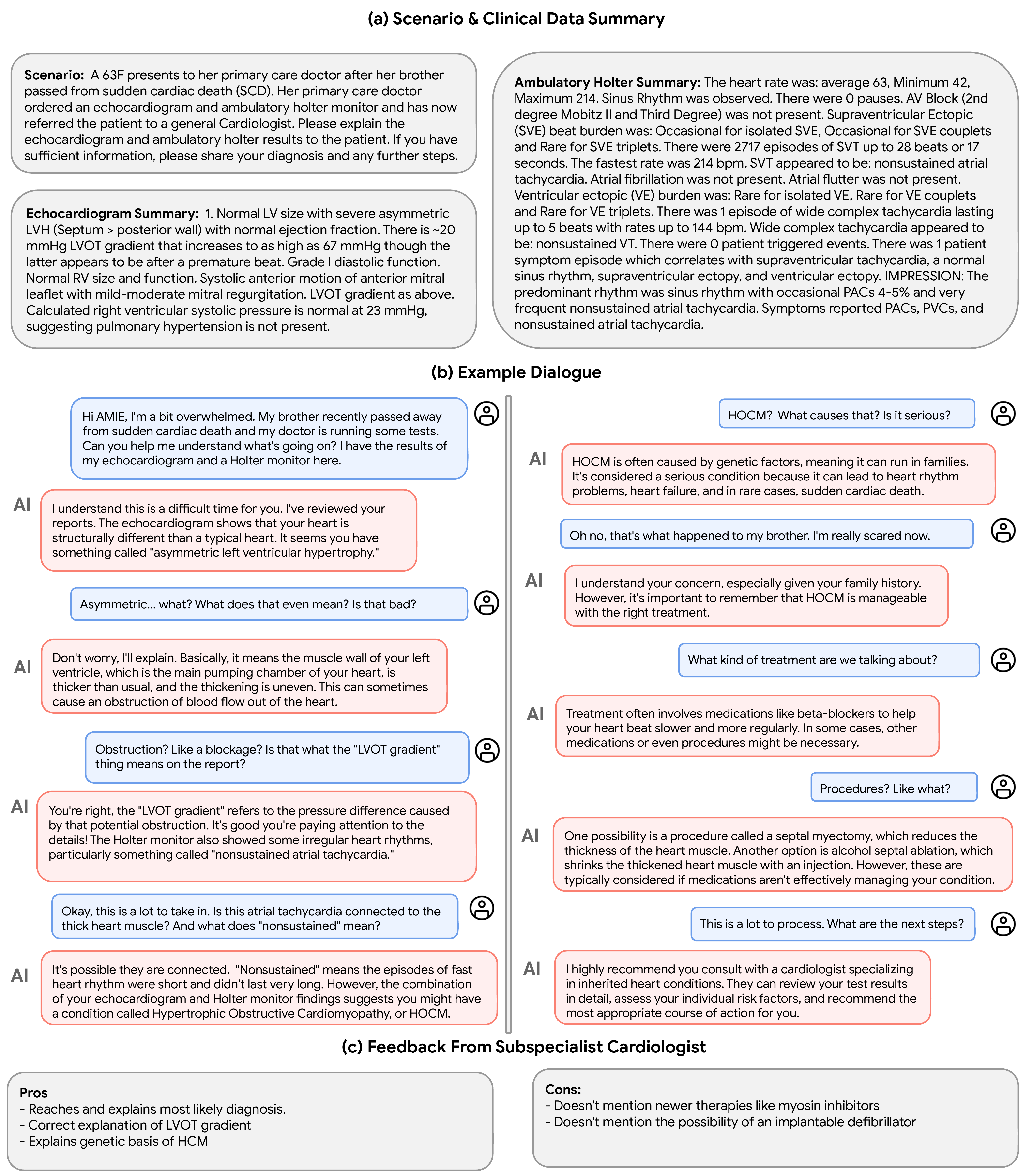}
    \vspace{0.2cm}
    \caption{\textbf{AMIE communicates with a patient about her diagnosis.} AMIE guides a patient through results of their echocardiogram and ambulatory holter monitor. The potential clinical application is AMIE as an always available tool to help patients with result interpretation.}%
    \label{fig:dialogue_pat10}%
\end{figure}

\begin{figure}[htp!]
    \centering
    \includegraphics[width=\textwidth]{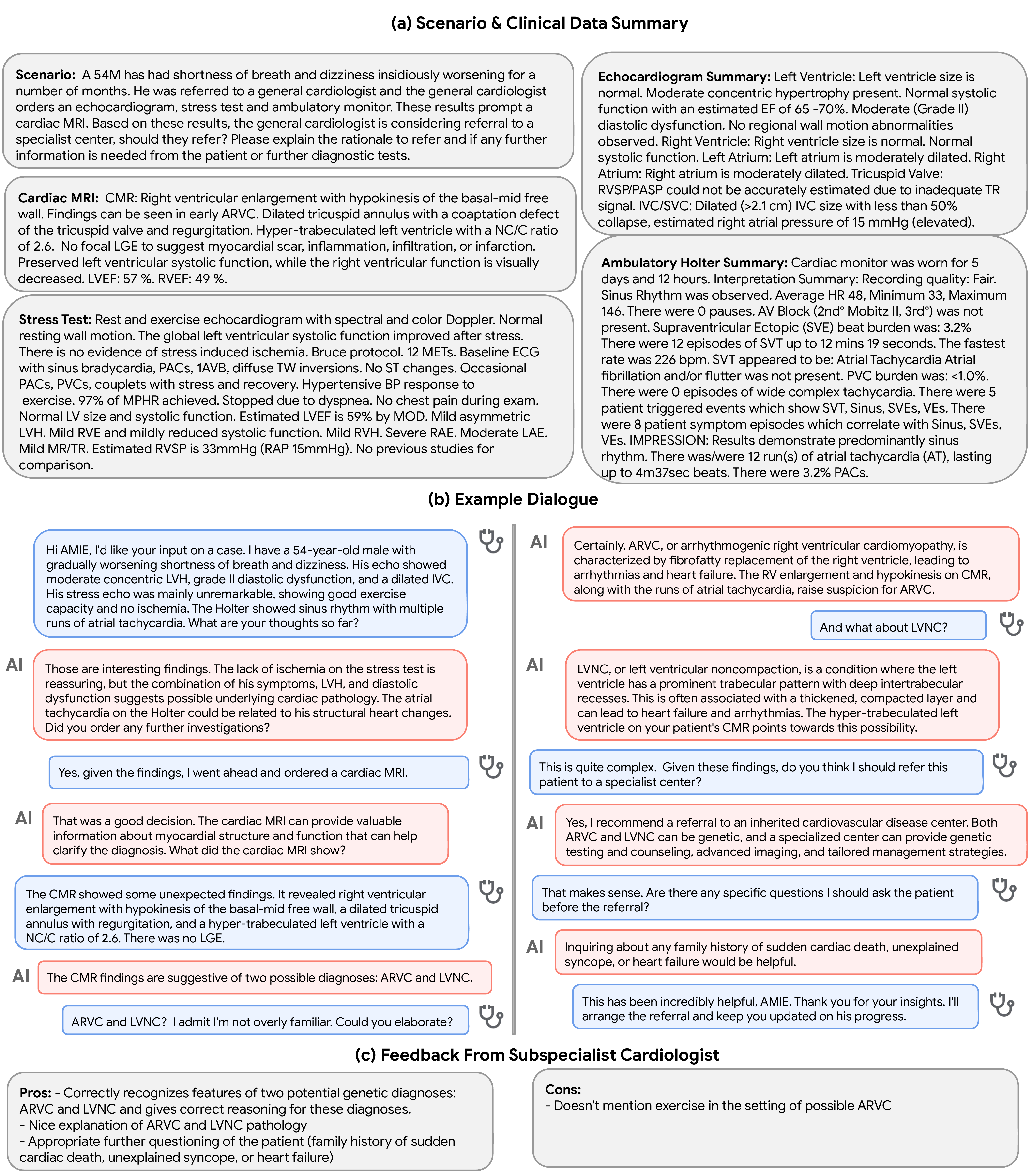}
    \vspace{0.2cm}
    \caption{\textbf{AMIE assists a general cardiologist.} AMIE discusses a patient with a general cardiologist and briefly explains rare genetic conditions ARVC and LVNC.}%
    \label{fig:dialogue_pat13}%
\end{figure}

\begin{figure}[htp!]
    \centering
    \includegraphics[width=\textwidth]{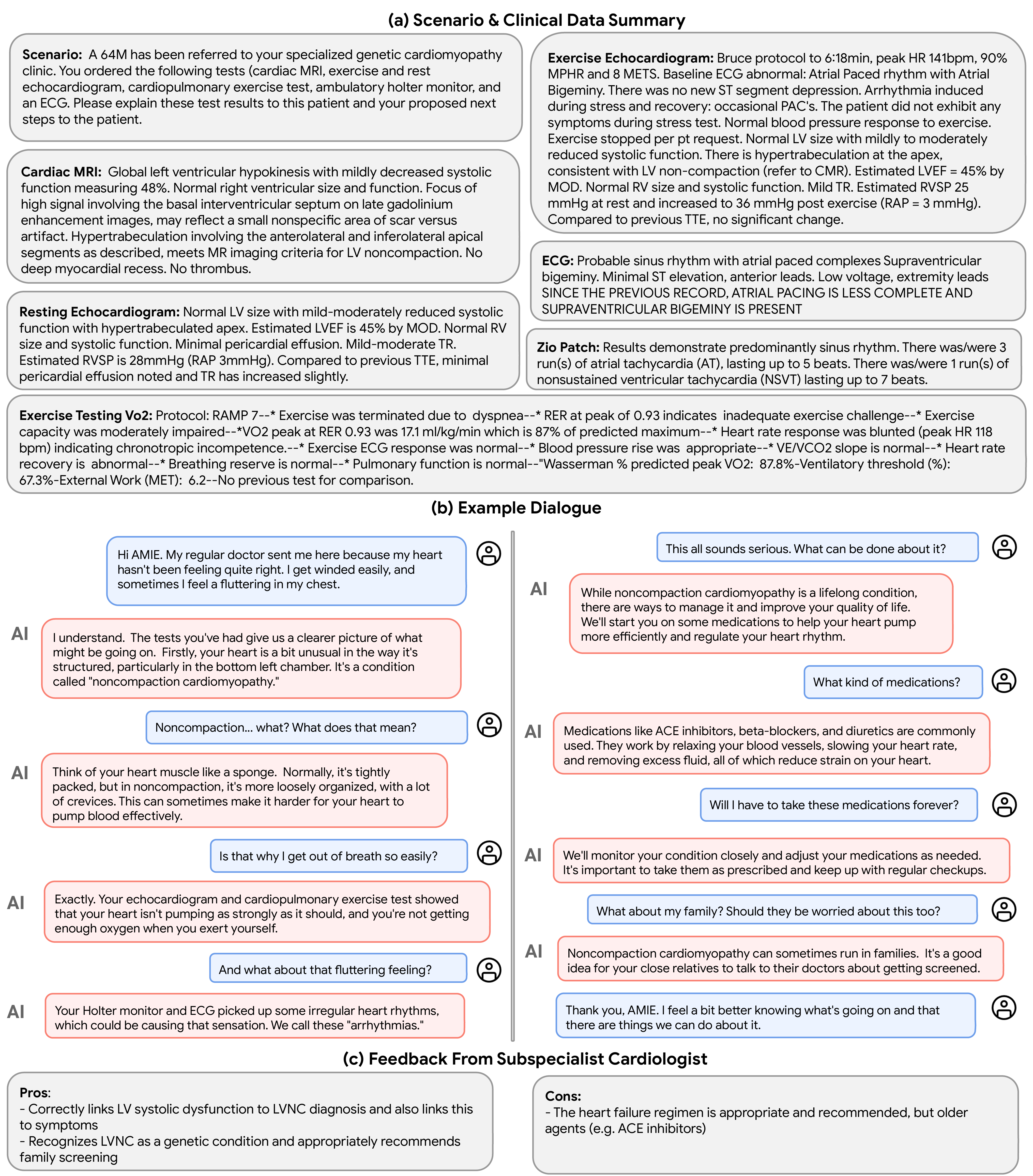}
    \vspace{0.2cm}
    \caption{\textbf{AMIE presents the patient assessment as a subspecialist.} AMIE acting as a subspecialist cardiologist delivers its assessment of a patient based on a wide range of testing.}%
    \label{fig:dialogue_pat2}%
\end{figure}

Here we present dialogue examples to illustrate potential clinical applications of AMIE in conveying information to patients and assisting general cardiologists in their assessment. The example scenarios used to generate these dialogues are described in \cref{tab:dialogue}, with summaries of the provided clinical data, resulting dialogues, and sub-specialist commentary for each scenario presented in \cref{fig:dialogue_pat10}, \cref{fig:dialogue_pat13}, and \cref{fig:dialogue_pat2}.

\begin{table}[htbp!]
\centering
\begin{tabular}{>{\centering\arraybackslash}p{8cm}>{\centering\arraybackslash}p{8cm}}
\toprule
Scenario & Explanation \\ \hline
\vspace{0.01cm}
A 63F presents to her primary care doctor after her brother passed from sudden cardiac death (SCD). Her primary care doctor ordered an echocardiogram and ambulatory holter monitor and has now referred the patient to a general Cardiologist. Please explain the echocardiogram and ambulatory holter results to the patient (review results for echocardiogram and holter monitor, without reviewing other test results). If you have sufficient information, please share your diagnosis and any further steps.  & \vspace{0.01cm} AMIE reaches a diagnosis and then explains it to a patient.\\ \hline
\vspace{0.01cm}
A 54M has had shortness of breath and dizziness insidiously worsening for a number of months. He was referred to a general cardiologist and the general cardiologist orders an echocardiogram, stress test and ambulatory monitor. These results prompt a cardiac MRI. Based on these results, the general cardiologist is considering referral to a specialist center, should they refer? Please explain the rationale to refer and if any further information is needed from the patient or further diagnostic tests. & \vspace{0.01cm} AMIE provides assistive dialogue for a general cardiologist. \\ \hline
\vspace{0.01cm}
A 64M has been referred to your specialized genetic cardiomyopathy clinic. You ordered the following tests (cardiac MRI, exercise and rest echocardiogram, cardiopulmonary exercise test, ambulatory holter monitor, and an ECG). Please explain these test results to this patient and your proposed next steps to the patient. & \vspace{0.01cm} AMIE assumes the role of the specialist cardiologist and presents an assessment. \\ \bottomrule
\end{tabular}
\vspace{0.1cm}
\caption{\textbf{Prompts for AMIE's simulated dialogue across three scenarios:} (1) AMIE reaches a diagnosis and then explains it to a patient. (2) AMIE provides assistive dialogue for a general cardiologist. (3) AMIE assumes the role of the specialist cardiologist and presents an assessment.}
\label{tab:dialogue}
\end{table}

\clearpage

\setlength\bibitemsep{3pt}
\printbibliography
\balance
\clearpage
\end{refsection}

\end{document}